\documentclass[a4paper,UKenglish]{lipics} 

\usepackage{macrosNew}
\Bleck

\usepackage{microtype}
\usepackage{amsfonts}
\usepackage{amssymb}
\usepackage{amsmath} 
\usepackage{graphicx}
\usepackage{tikz-cd} 
\usepackage{enumitem} 

\title{
Compositional security  and collateral leakage}

\author[1]{N Bordenabe}
\author[1]{A McIver}
\author[2]{C Morgan}
\author[1]{T Rabehaja}
\affil[1] {Dept. Computing,
 Macquarie University, Sydney}
\affil[2]{Data61 \& UNSW,
  Sydney}
\authorrunning{Bordenabe, McIver, Morgan and Rabehaja} 

\keywords{Quantitative information flow, program semantics, secure refinement.}

\newcommand\DaleniusVar[2]{ {#1}^#2  }

\begin{document}

\maketitle

\begin{abstract}
In quantitative information flow we say that program $Q$ is ``at least as secure as'' $P$ just when the amount of secret information flowing from $Q$ is never more than flows from $P$, with of course a suitable quantification of ``flow''.  
This secure-refinement order $\Ref$ is \emph{compositional} just when $P{\Ref}Q$ implies ${\cal C}(P){\Ref}{\cal C}(Q)$ for any context ${\cal C}$, again with a suitable definition of ``context''.

Remarkable however is that leaks caused by executing $P,Q$ might not be limited to their declared variables: they might impact  correlated secrets in variables declared and initialised in some broader context to which $P,Q$ do not refer even implicitly. We call such  leaks \emph{collateral} because their effect is felt in domains of which (the programmers of) $P, Q$ might be wholly unaware: our inspiration is the ``Dalenius'' phenomenon for  statistical databases \cite{Dalenius:1977aa,Dwork:2006aa}. 

We show that a proper treatment of these collateral leaks is necessary for a compositional program semantics for read/write ``open'' programs. By adapting a recent Hidden-Markov denotational model for non-interference security \cite{McIver:2014ab,McIver:15}, so that it becomes ``collateral aware'', we give techniques and examples (e.g.\ public-key encryption) to show how collateral leakage can be calculated and then bounded in its severity.
\end{abstract}

\section{Introduction}\label{s1336}

The problem of information disclosure in the context of statistical databases was formulated by Dalenius as an ideal privacy goal: 
``Nothing about an individual should be learnable from the database that cannot be learned without access to the database''
\cite{Dalenius:1977aa}.  Later he  argued the infeasibility of such a strict goal; more recently Dwork \cite{Dwork:2006aa} addressed the same concern
demonstrating that 
whenever there is a (known)  \emph{correlation} between two pieces of information, anything learned about one piece implies that something might also be learned about the other.  In secure programming generally, i.e.\ not only read-only databases, this corresponds to leaking information about a secret ``high-level'' variable {\Pf X}, which then consequentially leaks information about a different high-level variable {\Pf Z} that \emph{does not appear in the program at all}, but is known via ``auxiliary information'' to be correlated with the initial value of {\Pf X}. Because of the generality of this programming-language perspective, we call this effect \label{g13928}\emph{collateral leakage}.
 
Here we study this broader phenomenon of collateral leakage
in the general setting of read/write programs operating as ``open systems''. Because  \cite{Alvim:2014aa} studied Dalenius leakage in terms of abstract channels (read-only programs), and \cite{McIver:15} explored the semantics of information flow for read/write programs operating as ``closed systems'' without external correlation, this work can be see as bringing the two together:
we compute general bounds on collateral leakage for read/write open systems, and we adapt our earlier fully abstract program semantics to treat it compositionally. 

The following ``password'' example helps to illustrate the issues.
\newpage 

\begin{figure}
 {\Pf\small
  \begin{tabular}{c}
   // Password X is initially uniformly distributed over ${\CalX} = \{{\Pf A},{\Pf B},{\Pf C}\}$. \\\\[-2ex]
   \begin{tabular}[t]{ll}
    \textsf{``Lax'' user} \\\hline\\[-2.5ex]
    X\From\ [A,B,C] & $^\ast$ \\
    leak [{\Pf X}$^+$,{\Pf X}$^-$] & $^\dagger$ \\
   \end{tabular}
   \hspace{16em}
   \begin{tabular}[t]{ll}
    \textsf{``Strict'' user} \\\hline\\[-2.5ex]
    X\From\ [{\Pf X}$^+$,{\Pf X}$^-$] & $^\$$ \\
    leak [{\Pf X}$^+$,{\Pf X}$^-$] & $^\dagger$ \\
   \end{tabular}\\\\
   \begin{tabular}{l@{~}l}
    $^\ast$    & {\it {\Pf[...]} is the uniform distribution over {\Pf\{...\}}; and {\Pf X\From} assigns to {\Pf X} from a distribution.} \\
    $^\$$      & {\it {\Pf X}$^+$ is the letter following {\Pf X} in $\CalX$ (wrapping around), and {\Pf X}$^-$ the preceding.} \\[.5ex]
    $^\dagger$ & {\it {\Pf leak [X$^+$,X$^-$]} makes a (fair) choice secretly between {\Pf X}$^+$ or {\Pf  X}$^-$,
                           then emits the value somehow:} \\
               & {\it it does not however indicate whether it chose {\Pf X}$^+$ or {\Pf  X}$^-$ to leak.} \\
               & {\it Note that the {\Pf X} referred to in the {\Pf leak} statement is the updated value, after the {\Pf X\From}~update.}
   \end{tabular}
  \end{tabular}
 }

\bigskip
{\small \textsf{Lax} may choose any new password, uniformly, including his current; but \textsf{Strict} must \emph{change} his password, again uniformly. In both cases the distribution of the new {\Pf X} is again uniform: for \textsf{Lax} it is independent of {\Pf X}'s initial value; but for \textsf{Strict} is is correlated.
Both users, in the second statement $\dagger$, suffer an ``over the shoulder'' attack against the new password.

These programs, although presented informally here, have a precise denotation in the model of \cite{McIver:15}: this example is not relying on informality.}
\caption{Updating a password}\label{f1407}
\end{figure}
\label{g14422}
\subsection{Changing a password: is it only ``fresh'', or actually ``different''?}\label{s0917}

Consider the situation in \Fig{f1407} with state space a set $\{\textsf{A},\textsf{B},\textsf{C}\}$ of just three possible one-letter passwords (to keep things simple); we assume the distribution of passwords is known, and (again for simplicity) we assume it's uniform. Now contrast two users' behaviours when forced to change their passwords: User \textsf{Lax} \C{makes a fresh, uniform choice of password; User \textsf{Strict} however is forced  to choose a different password, still uniformly but not the one he already has. In \Fig{f1407} both users then suffer an ``over the shoulder'' attack where the adversary glimpses a letter the user did \emph{not} type as he logged-in with his new password.}

Although it's common to consider leaks wrt.\ initial values, especially e.g.\ for channels that don't update any state anyway, in \Fig{f1407} instead it's very natural to focus on 
the \emph{final} state of these two programs, i.e.\ the new password: the old password is not so interesting to the shoulder-cruiser. This ``focus on final'' obtains for the theory as well, but for a different reason: our aim (in program semantics) is to integrate security ``correctness'' with (ordinary) functional correctness of programs, i.e.\ to treat the two within the same framework \cite{McIver:15}; and since functional correctness (and correctness comparisons, i.e.\ refinement \cite{Morgan:94}) is determined wrt.\ the final values a program produces, we should do the same for security correctness: \C{more is said about this in \Sec{s0908} and its associated \App{a1002}.}

\C{The example above was deliberately constructed so that} the two programs have the same final distribution: with our assumption of uniformly distributed passwords beforehand, in both cases (\textsf{Lax},\textsf{Strict}) the new password's distribution will be uniform too. Furthermore, the effect of the {\Pf leak} statement is also the same: \C{the adversary has the same knowledge about the new password in both cases --- he knows exactly one value that it is not.} So are these programs equivalent in terms of their functional- and information-flow behaviour?

The answer depends on the context of operation.  As closed systems with a single secret ${\Pf X}$, they are indeed equivalent when the initial distribution is uniform.%
\footnote{Complex examples can give equivalence for all initial distributions; but this simple case makes our point.}
But they are not equivalent if we consider correlations between the final value of {\Pf X}, the new password, and some \emph{other} variable, call it {\Pf Z}, even though {\Pf Z} is not mentioned in either program. That is because to any adversary aware of that correlation, the first program will leak information about {\Pf Z} but the second will not: for example, if it is known that each user has the same password for their Facebook ({\Pf X}) and Twitter ({\Pf Z}) accounts, then the over-the-shoulder attack against \textsf{Strict}'s new password leaks information about his (unchanged) Twitter password \emph{even though his program does not access Twitter at all}.
\footnote{It's interesting to reflect that many \textit{IT} departments might impose \textsf{Strict}'s policy; but it's \emph{his} Twitter account that is at risk, not \textsf{Lax}'s.}
This is essentially the Delanius scenario presented in a programming-language context where ${\Pf X}$ is the statistical database and the correlation with ${\Pf Z}$ is ``auxiliary information'' \cite{Dwork:2006aa} except that, unlike in the traditional presentation, ours here allows the ``database'' (the password) to be updated. In this programming context, we call {\Pf Z} a \emph{collateral variable},
\label{g135448}and information flow from it is \emph{collateral leakage}.

The Dalenius phenomenon, so well known in security research, is truly remarkable as collateral leakage in rigorous reasoning about programs\C{: what kind of algebra would invalidate an equality because of variables not occurring in either formula?} In \Fig{f1407} are two sequential programs whose treatment of \emph{all} variables they mention ({\Pf X}) is the same in terms of assignments to those variables and leaks about them; yet the two programs are not the same if collateral variables must be considered, variables the program does not even mention.

In a collateral-enabled context, the semantics of \textsf{Lax} and of \textsf{Strict} must differ.%
\footnote{A more extreme example of this is the two programs {\Pf X:= 0} and {\Pf leak X; X:= 0} where both programs simply set {\Pf X} to 0 and so an adversary, knowing the code, knows also that {\Pf X} will finally be 0. Yet the second program can reveal something about a collateral {\Pf Z}, while the first cannot.}

\subsection{Our contributions}\label{s1013N}\label{s1450N}
\newcounter{Ca}
\begin{enumerate}
\item We extend our channel-based results from \cite{Alvim:2014aa}, based effectively on read-only programs, to show that even read/write programs' collateral leakage can be bounded without knowing what the collateral variables might be (\Sec{s1221}), 
\item \label{i1012c} \C{and that the bounds can be used in practice, e.g.\ in cryptography (\Sec{ss1055}).}
\item\label{i1012e} We explain  the connection between collateral leakage and compositional semantics for secure programs: that the ``problems'' with collateral leakage, in both informal (Dalenius) and rigorous (program semantics) settings, are the same (\Sec{s1330}).
\item\label{i1012f} We sketch how to extend our treatment of abstract channels \cite{McIver:2014aa} and abstract Hidden Markov Models \cite{McIver:15} \C{to construct fully abstract collateral-aware program semantics} (\Sec{s1449}).
\Cf{Make sure we mention ``fully abstract'' \textbf{in \Sec{s1449}}.}
\item We justify for our strong secure-refinement order, by reducing it to the simpler Bayes vulnerability \cite{Smith:2009aa} tests in a collateral context (\Sec{s0908}).
\setcounter{Ca}{\theenumi}
\end{enumerate}

In  (\Sec{s1139}) we provide the background for our contributions.
Full proofs are in the appendix. Notation is introduced as needed, with a full glossary  for reference in \App{s1241}.

\section{\C{Abstract} information flow: channels, hypers, \C{vulnerability}}\label{s1139}

\subsection{Abstract semantics of channels, based on hyper-distributions}\label{s1540}
Given set $\CalX$ of inputs and $\CalY$ of observations, a \emph{channel} between $\CalX$ and $\CalY$ is a (stochastic) matrix whose $\CalX$-indexed rows sum to 1. We write the type of such channels/matrices as $\CalX{\MFun}\CalY$ and for $C\In\CalX{\MFun}\CalY$ its constituents are elements $C_{x,y}$ at row $x$ and column $y$ that gives the conditional probability of output $y$ from input $x$, the $x$'th row $C_{x,-}$ and the $y$'th column $C_{-,y}$. For any set $\CalS$ we write $\Dist\CalS$ for the set of discrete distributions on $\CalS$; thus for example any row $C_{x,-}$ of $C\In\CalX{\MFun}\CalY$ can be interpreted as an element of $\Dist\CalY$. Similarly the $\CalX{\times}\CalY$ matrix-subtype $\CalX{\MFunR}\CalY$ is co-stochastic, a channel from $\CalY$ to $\CalX$  with 1-summing columns. 

Following \cite{McIver:2014aa,Alvim:2012aa} \C{an \emph{abstract} channel is a function} $\Dist\CalX{\Fun}\Dist^2\CalX$ from a \emph{prior} distribution on $\CalX$, i.e.\ of type $\Dist\CalX$,
\label{g140140}\label{g140911}
to a distribution of distributions-on-$\CalX$, thus of type $\Dist(\Dist\CalX)$, equivalently $\Dist^2\CalX$. That latter is the type of \emph{hyper-distributions} on $\CalX$, or ``hypers'' for short. To do that, we form the joint distribution induced by the prior and the channel, but then  abstract from the observations' values $y\In\CalY$, retaining only the effect they have on our a-posteriori, i.e.\ Bayesian reasoning that revises the prior. The support of the hyper is the set of posterior distributions induced on the prior by each possible observed value $y\In\CalY$, and the probability assigned by the hyper to each such posterior, i.e.\ each element of its support, is the (marginal) probability associated with the observation $y$ that induced it.
In more detail --- given a channel $C\In\CalX{\MFun}\CalY$ and prior $\pi\In\CalX$, the resulting hyper $\Delta\In\Dist^2\CalX$ is found by: constructing the joint distribution $J{\In}\Dist(\CalX{\times}\CalY)$ given by $J_{x,y}\Defs \pi_xC_{x,y}$; then for each $y$ the column $J_{-,y}$ normalising%
\footnote{If several distinct $y$'s produce the same posterior, they are amalgamated; if there is $y$ with zero marginal probability, it and its (undefined) posterior are omitted.}
to give the posterior induced on $\pi$ by that $y$; and finally taking for the probability assigned to that posterior the normalising factor.
We write $\pi{\Apply}C$ for the joint distribution $J$ and $[-]$ for the abstraction from $y$, so that $\Delta=[\pi{\Apply}C]$ describes the whole hyper-construction procedure.
\label{g1421}

As an example, in \Fig{f1407} above the statement  \textsf{leak [{\Pf X}$^+$,{\Pf X}$^-$]} acts, as a channel, by taking the (initial, and uniform) prior $[\textsf{A},\textsf{B},\textsf{C}]$ \C{produced by the assignment statement just before} to the final hyper $[\,[\textsf{B},\textsf{C}],[\textsf{C},\textsf{A}],[\textsf{A},\textsf{B}]\,]$ \C{where each of the posterior distributions is uniform (over two values), and the hyper itself is a uniform distribution on the three of them. An adversary does not know beforehand which two values afterwards he will know are possible; afterwards he does know the two possibiities, but still does not know which is right.%
\footnote{The first is an ``unknown known'' and the second is a ``known unknown''. }
}

\subsection{\C{Vulnerabliity induced by gain-functions}}\label{s1216}
\C{Vulnerability \cite{Alvim:2012aa} is} a generalisation of entropy (of distributions), no longer necessarily e.g.\ Shannon but now others more adapted  for secure programming, and whose great variety allows fine-grained control of the significance of the information that might be leaked \cite{Alvim:2012aa,Alvim:14}.

Given a state-space $\CalX$, \C{vulnerability is induced by a} \emph{gain function} over that space, typically $g$ of type $\GainF{\CalW}\CalX = \CalW{\Fun}\CalX{\Fun}\Real$, for some space of \emph{choices} $w\In\CalW$. When $\CalW$ is obvious from context, or unimportant, we will omit it and write just $g\In\GainF{}\CalX$.
\Cf{I'm not restricting to $[0,1]$ for the moment. {\Ax we don't need to restrict to $[0,1]$ but we do need non-negative reals?\CBar}}
Given $g$ and $w$ (but not yet $x$) the function $g.w$ is of type $\CalX{\Fun}\Real$
\footnote{We write dot for function application, left associative, so that function $g$ applied to argument $w$ is $g.w$ and then $g.w.x$ is $(g.w)$ applied to $x$, that is using the Currying technique of functional programming. This convention reduces clutter of parentheses, as we see later.}
and can thus be regarded as a random variable on $\CalX$. As such, it has an expected value on any distribution $\pi$ over $\CalX$, written $\Exp{\pi} g.w$.%
\footnote{In general we write $\Exp{\pi}{f}$ for the expected value of function $f\In\CalX{\Fun}\Real$ on distribution $\pi\In\Dist X$.}
\label{g1416}

Once we have $x$, the (scalar) value $g.w.x$ is simply of type $\Real$ and represents the gain to an adversary if he guesses $w$ when the secret's actual value is $x$. This suggests that $\CalW$ should simply be $\CalX$, i.e.\ that there would be no point in guessing a $w$ that was not in $\CalX$, and indeed a particularly simple example is $\CalW{=}\CalX$ with $g.w.x = (\textrm{1 if $w{=}x$ else 0})$ so that the adversary gains 1 if he guesses correctly and 0 otherwise: we call this particular gain-function $\BVg$. 
\label{g1249} However for reasons explored thoroughly elsewhere \cite{Alvim:2012aa,McIver:2014aa}, there are  \emph{practical} advantages to allowing $\CalW$ to be more general.
And a \emph{theoretical} benefit is that it is the more general $\CalW$'s that allow representation of many conventional entropy functions (including even Shannon, after some encoding), thus bringing them all within the same framework \cite{McIver:2014aa}.

\label{g1413}A gain function $g\In\GainF{}\CalX$ induces a \emph{$g$-vulnerability} function $V_g\In\Dist\CalX{\Fun}\Real$ so that $V_g[\pi]$ for $\Pi\In\Dist\CalX$ is the maximum over all choices $w\In\CalW$ of the expected value of $g.w$ on $\pi$, that is $\max_w(\Exp{\pi}{g.w})$. In the simple 1-or-0 case above, the vulnerability $V_\BVg$ is called the \emph{Bayes vulnerability}, sometimes written just $V$; it is one-minus the Bayes-Risk of Decision Theory.
\label{g1423}

{\Cx Vulnerability applies also to hypers, so that $V_g\Delta$ for $\Delta\In\Dist^2\CalX$ is the expected value of $V_g$ over $\Delta$, itself considered as a distribution on $\Dist\CalX$. That allows us to write the succinct $V_g[\pi{\Apply}C]$ for the a posteriori $g$-vulnerability of prior $\pi$ through channel $C$.%
\footnote{Recalling that $[-]$ around a list of one element makes the point distribution on that element, we can regard $V_g[\pi]$ as the expected value of $V_g$ on the point hyper on $\pi$, agreeing with our usage just above.}
}

\subsection{\C{(Classical) Gain-function leakage and capacity for channels}}\label{s1003}

In general, information leakage on prior $\pi$ due to channel $C$ is a comparison between the ``information content'' of prior knowledge about $\pi$, and of posterior knowledge  after observing $C$'s output; in particular, its precise definition depends on how the information is measured. Here we will use gain functions and speak of $g$-leakage (rather than e.g.\ the more specific Shannon leakage).
\label{g1428}
The \emph{multiplicative $g$-leakage} of $C$ wrt prior $\pi$ and gain function $g$ is the ($\log_2$ of the) ratio between the posterior and prior $g$-vulnerabilities:
\begin{equation}\label{e1244}
 \call_g(\pi, C)\,\Defs\quad
 \lg (\,{\Tx V_g[\pi{\Apply}C]}\,/\,V_g[\pi]\,)
 \quad.
\end{equation}
The \emph{capacity} of a channel is the supremum of that leakage \Eqn{e1244}, \C{but varying in its definition depending on whether the supremum is} over either
gain functions,  or priors or both:
\[
 \call_\forall(\pi, C)\,\Defs\sup_g \call_g(\pi, C)
 \quad,\quad
 \call_g(\forall, C)\,\Defs\sup_\pi \call_g(\pi, C)
 \quad,\quad
 \call_\forall(\forall, C)\,\Defs\sup_{\pi,g} \call_g(\pi, C)
 \quad.
\]
Remarkably, it can be shown that  $\call_\forall(\forall, C)$ equals $\call_\BVg(\forall, C)$ (``min-capacity''):
it is the most robust estimation of leakage, and can always be achieved for some scenario and  prior \cite{Alvim:2012aa}.

\subsection{\C{\emph{Collateral} Gain-function leakage and capacity for channels}}\label{s1124}
The \emph{\C{$\CalZ$-}collateral leakage} of a channel $C\In\CalX{\MFun}\CalY$ is with respect to some third space $\CalZ$ and is induced by knowledge of a correlation between values in $\CalX$ and $\CalZ$.%
\footnote{In Dalenius terms $\CalX$ is the statistical database, with queries of type $\CalY$, and the correlation between $\CalX$ and $\CalZ$ is auxiliary information \cite{Dwork:2006aa}.}
We write {\Pf X},{\Pf Y},{\Pf Z} of resp.\ types $\CalX,\CalY,\CalZ$ equivalently for random- or program variables, depending on context.
We assume a known correlation between {\Pf X} and a fresh secret {\Pf Z} given by a joint probability distribution $\Pi\In\Dist(\CalZ{\times}\CalX)$ so that leaks about {\Pf X} through (a computer program implementing) channel $C$ induce ``collateral'' leaks about {\Pf Z} as follows. 
\label{g1430}
Define $\leftmarg\pi$ to be the $\CalZ$-marginal of $\Pi$ and determine some channel $\rightchan\Pi$ in $\CalZ{\MFun}\CalX$ so that $\Pi = \leftmarg\pi{\Apply}\rightchan\Pi$: this factors $\Pi$ into its marginal and a conditional.%
\footnote{Recall that $\Pi_{z,x}$ is thus $\leftmarg\pi_z\rightchan\Pi_{z,x}$. Unless marginal $\leftmarg\pi$ is full support, conditional $\rightchan\Pi$ is not unique. But it does not matter: any choice suffices, and does not affect the outcome.}
Then the matrix multiplication $\rightchan\Pi{\MMult}C$ gives a channel $D$ (for Delanius) of type $\CalZ{\MFun}\CalY$, and the collateral \emph{leakage} of $C$ is the ordinary leakage of $D$, calculated as at \Eqn{e1244} above. (The {\Ax (matrix)} multiplication $\rightchan\Pi{\MMult}C$ is the \emph{channel cascade} of $\rightchan\Pi$ and $C$.)

The $\CalZ/\Pi$-collateral \emph{capacity} of channel $C$ is then the upper bound of the above with respect to all gain functions, that is $\call_\forall(\leftmarg\pi, \rightchan\Pi{\MMult}C)$. {\Cx The collateral capacity of channel $C$ in general is then taken over all possible $\CalZ$ and $\Pi$.}
\Cf{Do we return to this somewhere below? Do we cite \Cite{CSF14}?}

\section{\C{Collateral gain-function leakage and capacity for \HMM's}}\label{s1221}
\subsection{\HMM's as models for leaking probabilistic programs}\label{s1735}
\C{The discussion in \Sec{s1139}, for channels only, corresponds to considering read-only programs.}
With \emph{Hidden Markov Models} we combine channels and  Markov transitions between initial- and final states (e.g.\ over variables {\Pf X} in $\CalX$), \C{thus modelling read/write programs}; and the constructions of \Sec{s1139} carry over. A Markov transition  is described by a matrix $M$ say so that $M_{x,x'}$ is the probability that initial state $x$ will result in final state $x'$. An \emph{\HMM-step} then comprises a channel and a transition together, but acting independently on the initial state: we call $C$ its channel and $M$ \emph{markov} (lower case), and write it $\CM{C}{M}$ of type $\CalX\MFun\CalY{\times}\CalX$. Defined $\CM{C}{M}_{x,y,x'} = C_{x,y}M_{x,x'}$, it is a single stochastic matrix with rows $\CalX$ and columns $\CalY{\times}\CalX$.

\HMM-steps have the special characteristic that their $C$ and $M$ effects (i.e.
 output values $y,x'$) are independent (uncorrelated) for each separate input-value $x$; but two steps, say $H^{1,2}$ of the same type $\CalX\MFun\CalY{\times}\CalX$ can be \emph{sequentially composed} to give a single \HMM\ again: \C{this is natural if we are using them to model programs. The composed type} is the $\CalX\MFun\CalY^2{\times}\CalX$ that takes initial state $x$ to final state $x'$ via some intermediate state $x''$, leaking information $(y^1,y^2)$ --as it goes-- gradually into the set $\CalY^2$. We define
 \label{g1434}
\begin{equation}\label{e1555}
 (H^1;H^2)_{x,(y^1,y^2),x'}
 \Wide{=}
 \sum_{x''} H^1_{x,y^1,x''}H^2_{x'',y^2,x'} ~,
\end{equation}
and note that it is again stochastic, \C{but no longer necessarily a \emph{step} --- its} 
outputs $y^2$ and $x'$ might be correlated even for a single input $x$.
\Cf{Explain this better: do it in the appendix, in the step-section?}
Because of that, sequential compositions are strictly more general than the \HMM-steps built directly from $\CM{C}{M}$ --- that is, for $C^{1,2},M^{1,2}$ in general there is not necessarily a single $C,M$ such that $\CM{C^1}{M^1};\CM{C^2}{M^2} = \CM{C}{M}$. In \App{a1005} we give some elementary properties of \HMM-steps, showing in particular how the different encodings of markovs and channels preserve their different purposes and allows the same definition of sequential composition to be used for both.

More generally, two \HMM's over the same state $\CalX$ but distinct observation spaces $\CalY^{1,2}$ can be composed as just above; in that case the composite observation space is $\CalY^1{\times}\CalY^2$, and that is what allows us to use them for modelling programs which, in general, are not simply just \C{a fixed sequence of assignment statements}. {\Cx A monadic account of this generality, including loops and conditionals,
 is given in \cite{McIver:2014ab}.}

\subsection{The induced channel of an \HMM}\label{s1101}
 Recall that an \HMM\ of type $\MH$ takes an initial state in $\CalX$ to a final state (also) in $\CalX$ and leaks observations in $\CalY$ along the way, where \C{``along the way'' is more general than channels:  in \Fig{f1407} for example the {\Pf leak} statement occurred \emph{after} a Markov update.} Collateral leakage however, as we saw in the previous section, is with respect to a value {\Pf Z} which the \HMM\ (or program) does not even mention: it neither reads nor updates it; the programmer has never even heard of it. But it is somehow correlated with the initial state: that being so, we \C{show how to isolate} the channel part of an \HMM, \C{combining all its leaks and ignoring its final state.} Fix $H$ as some \HMM\ of type $\MH$ below.

\begin{definition}\label{d1429}
The \emph{effective channel} of $H$, written $\chan.H$, is a stochastic matrix of type $\CalX{\MFun}\CalY$ and defined simply by ignoring the final state: thus $(\chan.H)_{x,y}\Defs \sum_{x'}H_{xyx'}$~. 
\end{definition}
\begin{definition}\label{d1450}
The collateral leakage resp.\ capacity of $H$ wrt a prior $\Pi\In\Dist(\CalZ{\times}\CalX)$ is the collateral leakage resp.\ capacity of 
$\rightchan\Pi{\MMult}\chan.H$. \C{(This is well-defined, though $\rightchan\Pi$ is not unique.)}
\end{definition}

The simplicity of Defns.~\ref{d1429},\ref{d1450} conceals that it can be difficult in practice to calculate the collateral leakage of an \HMM, since \Def{d1429} mandates calculating the whole $H$ first, only then abstracting from its final state. Yet if $H$ is expressed as a sequential composition of many smaller ones, e.g.\ we have $H = H^1;H^2;\cdots;H^N$, still the final states of the \emph{intermediate} $H^n$'s must be retained, not only to form the composition, but because the overall $y$ observation from $H$ comprises all the smaller observations $y^1\cdots y^N$ with each $y^{n{+}1}$ being determined by the final state ${(x')}^n$ of the $H^n$ just before --- we can abstract only at the very end.

In the special case however where each $H^n$ is an \HMM-step $\CM{C^n}{M^n}$, the calculation of the effective channel can be somewhat decomposed.
\begin{lemma}\label{l1029} 
Let $H$ be an \HMM\ and $\CM{C}{M}$ an \HMM-step. Then \AppFrom{proof in \App{a1626}}
\[
 \begin{array}{lllp{10em}}
  \chan.\,\CM{C}{M}   &=& C \\
  \chan.\,(\CM{C}{M};H) &=& C\Par(M{\MMult}\chan.H)
\end{array}
\]
where \C{in general $(C^1{\Par}C^2)_{x,(y^1,y^2)} = C^1_{x,y^1}C^2_{x,y^2}$} is parallel composition of channels. The $M$ cannot be discarded, since it affects the prior of the ``tail'' $H$ of the sequential composition.
\end{lemma}

Even with \Lem{l1029}, in general $\chan.H$ can be challenging to compute because its size (given by the number of columns in the stochastic matrix representation) grows exponentially with the number of single-step \HMM's, in the definition of $H$, that have non-trivial channel portions.
We give an example of such a calculation in \Sec{ss1055} (fast exponentiation for cryptography).

On the other hand, if we want to compute only the collateral \emph{capacity}, we can obtain at least an upper bound at considerably less cost,  without the need to compute $\chan.H$ exactly. The following  provides an upper bound for $\call_\forall(\forall, \chan.H)$,  and requires only linear resources.

\begin{lemma}\label{l1519} For any $H$ let $\CCap.H$ be defined
 \AppFrom{proof in \App{a1626}} \\
\[
 \begin{array}{lll@{\hspace{2em}}r}
 \CCap.\,\CM{C}{M}     &=& {\Ax \call_\forall(\forall, C)} & \textrm{if $H{=}\CM{C}{M}$} \\
  \CCap.\,(\CM{C}{M};H') &=& {\Ax \call_\forall(\forall,C) + \min(\call_\forall(\forall, M), \CCap.H')}
    & \textrm{if $H{=}\CM{C}{M};H'$}
\end{array}
\]
Then $\call_\forall(\forall, \chan.H) \le \CCap.H$ with the stochastic matrix $M$ (on rhs) treated as a channel.
\end{lemma}

In fact \Lem{l1519} provides a very robust estimate of the collateral capacity of an \HMM, since it does not mention $\CalZ$ or the correlating $\Pi$.  And it is  the best possible general bound, achieving equality for some examples, 
e.g.\ \Fig{f1407}. {\Ax It is also easy to calculate since for any channel we have from \cite{Alvim:2012aa} that $\call_\forall(\forall,C)  = \call_\BVg(\Uni{\CalX}, C)$, where ${\Uni{\CalX}}$ is the uniform prior on ${\cal X}$. \label{g1519}}
On the other hand, the bound can be very conservative; we provide an example in \App{a1626}.

In cases where we know the correlation $\Pi$ (and thus $\CalZ$), we can compute the collateral capacity by identifying the gain function that achieves the \C{extremal} value in \Def{d1450}. 
\begin{theorem}
\label{t1008}
Given $H$ and $\Pi\In\Dist(\CalZ{\times}\CalX)$ with $\leftmarg\pi,\rightmarg\pi$ resp.\  the marginals \AppFrom{proof in \App{a1626}} \\
of \/ $\Pi$ on $\CalZ,\CalX$; define conditional $\rightchan\Pi$ as in \Sec{s1124}. There exists $\hat{g}\In\GainF{\CalZ}\CalZ$ and $\hat{g}^{\joint}\In \GainF{\CalZ}\CalX$ such that
\[
\call_{\forall}(\leftmarg\pi, \rightchan\Pi{\MMult} \chan.H) \Wide{=} \call_{\hat{g}}(\leftmarg\pi, \rightchan\Pi{\MMult} \chan.H) \Wide{=} \call_{\hat{g}^{\joint}} (\rightmarg\pi, \chan.H)~.
\]
\end{theorem}
This shows that it is possible to construct the gain-function
that maximizes the collateral capacity, and allows its exact calculation.
Moreover, it also shows that the collateral capacity of $H$ wrt.\ 
$\CalZ$ can be understood as regular $g$-leakage of $H$ wrt.\
the \emph{initial} state of $\CalX$.



The next theorem is more general, and gives an upper bound over all possible correlations: it is determined by the \C{extremal} leakage of the initial prior $\pi\In\Dist\CalX$, {\Ax thus easy to calculate \cite{Alvim:2014aa}.}

\begin{theorem}\label{t1203}
Given $H$ and $\Pi$ as above, $\Uni{\CalX}$ is uniform on ${\cal X}$,
then \AppFrom{proof in \App{a1626}}
\[
\call_\forall(\leftmarg\pi, \chanz{\MMult}\chan.H)
\Wide{\le}
\call_\BVg(\Uni{\CalX}, \chan.H)~, ~~~~\textit{where $\leftmarg\pi,\rightchan\Pi$ are as defined in \Thm{t1008}.}
\]
\end{theorem}
Note that when {\Pf X},{\Pf Z} 
are completely correlated, i.e.\ when $\leftchan\Pi$ and $\rightchan\Pi$ are both the identity, the inequality in \Thm{t1203} becomes equality. 

\section{Example: Collateral damage from leaky cryptography}\label{ss1055}
Keys for public-key cryptography are best if independent; but recently  \cite{Lenstra:12} discovered an unexpected sharing of the prime numbers used to generate them. Thus information leaked while using one secret key to encrypt a message, even if the message itself remains secure, could cause collateral leakage wrt.\ some other key elsewhere and so put future encryptions at risk --- even if they are at another site apparently having no connection to the first. That motivates our example here, the collateral leakage from a fast-exponentiation algorithm.

\begin{figure}
{\Pf\begin{tabular}{ll}
\multicolumn{2}{l}{\Cx// B for base, the cleartext; E for exponent, the key:\ precondition is B,E >= 0,0 .} \\
\multicolumn{2}{l}{\Cx// P for power, the ciphertext.} \\
P:= 1 \\
\multicolumn{2}{l}{while E!=0\quad \Cx// Invariant is P*(B\textasciicircum E) = $b^e$, where $b,e$ are initial values of B,E .} \\
~~ D\From\ [2,3,5] & // D for divisor;\ uniform choice from \{2,3,5\}. \\
~~ R:= E mod D; & // R for remainder.\\
~~ if R!=0 then P:= P*B\textasciicircum R fi & // \fbox{Side-channel}:\ is E divisible exactly by D ? \\
~~ B:= B\textasciicircum D & // D is small:\ assume no side-channel here. \\
~~ E:= E div D & // State update of E here.\ (No side-channel.) \\
end \\
\multicolumn{2}{l}{// Now P=$b^e$ and E=0:\ but what has an adversary learned about the initial $e$~?}
\end{tabular}}

\bigskip
{\small Although our state comprises {\Pf B},{\Pf E},{\Pf P},{\Pf D},{\Pf R} we concentrate only on the secrecy of {\Pf E}. \C{In particular, we are not trying to discover {\Pf B} or {\Pf P} in this case; and {\Pf D},{\Pf R} are of no external significance afterwards anyway.}}
\caption{Defence against side channel analysis in exponentiation}\label{f1143}
\end{figure}
\Fig{f1143} implements fast exponentiation with a random choice of divisor that defends against a side channel that leaks program flow (of a conditional) \cite{Walter02a}. Since the program code is public, that leak is effectively of whether divisor {\Pf D} exactly divides the current value of {\Pf E}, which value is steadily decreased by the update at the end of each iteration: thus additional information is leaked every time. In the standard (and fastest) algorithm the divisor {\Pf D} is always $2$, but that ultimately leaks {\Pf E}'s initial value exactly, one bit (literally) on each iteration. The final value of {\Pf E} is always zero, of no significance; but its initial value represents collateral leakage about subsequent use of this same key (obviously), but also the use of other apparently unrelated keys elsewhere \cite{Lenstra:12}. The obfuscating defence is to vary the choice of {\Pf D} unpredictably from one iteration to the next, choosing it secretly from some set ${\cal D}$, here $\{2, 3, 5\}$ although other options are possible. The divisor {\Pf D} itself is not leaked.

We modelled the loop as a \C{sequential composition of \HMM-steps for a fixed number of iterations} 
\N{
and used \Lem{l1029} to construct a channel that captures the leakage of information about the initial state of the program. 
\C{Although our calculation is wrt.\ the uniform prior on {\Pf E}, and even though we do not know the extent of any correlation between this key {\Pf E} and others used elsewhere,}
by using the ``Miracle Theorem'' \cite[Thm 5.1]{Alvim:2012aa}, we can
bound the maximum leakage about the initial value of {\Pf E} with the min-capacity
of such a channel. 
Furthermore, by relying on \Thm{t1203} we can see that this min-capacity can also be used as a bound on the collateral leakage \emph{with respect to any other secret that might be correlated to {\Pf E}}.
}

Detailed \C{calculations} are in \App{s1157Z}. Our \Tbl{table:leakage}
confirms that the larger the divisor set $\CalD$, the less effective is the side channel; and the protection is increased with \C{more bits for} {\Pf E}.
\begin{table}
\centering
\begin{tabular}{cc}
\begin{tabular}[t]{|c|c|c|c|}
\hline
Size of ${\Pf E}$ &$\CalD{=}\{2\}$ &$\CalD{=}\{2,3\}$&$\CalD{=}\{2,3,5\}$\\
\hline
4 bits &4&2.80& 2.22\\
5 bits &5&3.32& 2.61 \\
6 bits &6&3.83& 2.92\\
7 bits &7&4.34& 3.21 \\
8 bits &8&4.88& 3.51\\
\hline
\end{tabular}
&
\raisebox{-.6em}{\parbox[t]{.3\textwidth}{
Note that in the case $\CalD{=}\{2\}$ the whole secret {\Pf E} is leaked.
\medskip

\C{As explained at end \Sec{s1003}, that $\call_\BVg$ gives the upper bound $\call_{\forall}$ for all vulnerabilities.}
}}
\end{tabular}
\medskip
\caption{Collateral leakage $\call_\BVg(\forall, \textit{Prog})$ in bits wrt {\Pf E}  for different $\CalD$'s, for \textit{Prog} given at \Fig{f1143}. } \label{b1754}\label{table:leakage}
\end{table}
%
%

\section{The \C{abstract semantics} of ``collateral aware'' \HMM's}\label{s1138}
\subsection{Bits-leaked vs.\ a partial order on information-flow}\label{s1330}
We now contrast our practical example in \Sec{ss1055}, just above, with our earlier treatment of passwords, in particular our description (at the end of \Sec{s1540}) of leakage abstractly as a hyper-distribution rather than as a number of bits (\Tbl{table:leakage}).

Both password programs take prior $[\textsf{A},\textsf{B},\textsf{C}]$ to the hyper $[\,[\textsf{B},\textsf{C}],[\textsf{C},\textsf{A}],[\textsf{A},\textsf{B}]\,]$ on the final state,
\marginpar{$\dag$}
as we saw in \Sec{s1540}. But for collateral leakage we must consider (also) the initial state: and here the two programs differ. The hyper representing our knowledge of the prior distribution of the \emph{initial} password, after execution of \textsf{Lax}, is still $[\,[\textsf{A},\textsf{B},\textsf{C}]\,]$ --- it expresses that we know for sure (point hyper $[\,[{\cdots}]\,]$) that the distribution of {\Pf X} is uniform ($[\textsf{A},\textsf{B},\textsf{C}]$), that we have learned nothing (more) about the distribution of the password initially. But for program \textsf{Strict}
the hyper describing our revised knowledge about the initial password is
\begin{equation}\label{e0906}
 [\quad[\textsf{A}\At{\NF{1}{2}},\textsf{B}\At{\NF{1}{4}},\textsf{C}\At{\NF{1}{4}}],\quad
    [\textsf{B}\At{\NF{1}{2}},\textsf{C}\At{\NF{1}{4}},\textsf{A}\At{\NF{1}{4}}],\quad
    [\textsf{C}\At{\NF{1}{2}},\textsf{A}\At{\NF{1}{4}},\textsf{B}\At{\NF{1}{4}}]\quad] \quad,
\end{equation}
\label{g1426}%
now a uniform hyper over three, different, \emph{skewed} inners: the first $[\textsf{A}\At{\NF{1}{2}},\textsf{B}\At{\NF{1}{4}},\textsf{C}\At{\NF{1}{4}}]$, for example, is the distribution \textsf{A} wprob (with probability) 
\label{g1251}
$\NF{1}{2}$ and \textsf{B},\textsf{C} wprob $\NF{1}{4}$ each.%
\footnote{The details of calculating this hyper are given in \App{a1105}. Intuitively, whichever (final) not-password is seen by the adversary, it is more than uniformly probable that the initial password was that since, by \textsf{Strict}'s policy, the old password cannot be the new password that the adversary did not see.}
An adversary, knowing the program text of \textsf{Strict} and having calculated this hyper would conclude that, with probability $\NF{1}{3}$ each, he can after the run of the program revise his knowledge of the initial-password distributions from uniform to one of these three skewed inners (although he does not know beforehand which one of those three possible revisions will eventuate).

\emph{Abstracting} from any particular vulnerability (or entropy), we conclude that although \textsf{Lax} and \textsf{Strict} are equivalent wrt their setting and security of the new password when the previous password was uniformly distributed (\Sec{s0917}), wrt collateral security \textsf{Strict} is less secure than \textsf{Lax}. This abstract conclusion, i.e.\ reached without calculating specifics of ``number of bits leaked'', comes from the ``no less secure than'' partial order ($\Ref$) on hypers $\Dist^2\CalX$ where indeed
\(
 \textrm{\Eqn{e0906}}
 \SRef
 [\,[\textsf{A},\textsf{B},\textsf{C}]\,],
\)
that is the \textit{lhs}-hyper is strictly less secure than the \textit{rhs}-hyper
\cite{mcivermeinicke10a,Alvim:2012aa,McIver:2014ab,Alvim:2014aa,McIver:2014aa,McIver:15}. 
\label{g1447}
The significance is that the theory associated with ($\Ref$) shows that for \emph{any} hypers $\Delta^{1,2}$ we have $\Delta^1\Ref\Delta^2$ exactly when $V_g(\Delta^1)\geq V_g(\Delta^2)$ for \emph{all} gain-functions $g$: increased security corresponds to decreased vulnerability \cite{Alvim:2014aa}.
Since gain functions can express not only Bayes Vulnerability but also Shannon entropy (with some encoding) and many other entropies besides, that is a very strong result: for any ``favourite entropy'' you might use to compare the security of two programs, if it is expressible as $V_g$ for some $g$ (no matter how bizarre) then you will find the former at least as secure as the latter.

For example, the Shannon entropy of $[\,[\textsf{A},\textsf{B},\textsf{C}]\,]$ is $\lg3\approx1.58$; of \Eqn{e0906} it is $\NF{1}{2}{\cdot}1+\NF{1}{4}{\cdot}2+\NF{1}{2}{\cdot}2 = 1.5$. (Decreasing Shannon entropy indicates increasing vulnerability.) And the Bayes Vulnerability of $[\,[\textsf{A},\textsf{B},\textsf{C}]\,]$ is $\NF{1}{3}$; of \Eqn{e0906} it is $\NF{2}{3}$, increasing as expected.

\subsection{Collateral-aware fully-abstract semantics of \HMM's}\label{s1449}
All proofs for this section are to be found in \App{a1215}.

We saw in \Sec{s1540} an abstract semantics for (read only) \emph{channels} in $\Dist\CalX{\Fun}\Dist^2\CalX$; we could summarise that by writing $\HMMSem{C}.\pi = [\pi{\Apply}C]$, that is that $\HMMSem{-}$ is the semantic function that takes a channel matrix in $\CalX\MFun\CalY{\times}\CalX$ to a function as just above. For \HMM's we do similar:
\begin{definition}\label{d1207}
Given $H\In\MH$ its denotation in $\Dist\CalX{\MFun}\Dist^2(\CalX{\times}\CalX')$ takes prior $\pi$ to the hyper $[\pi{\Apply}H]$ where for hypers (generalising channels) we define $(\pi{\Apply}H)_{x,y,x'} = \pi_x H_{x,y,x'}$, and using $\CalY$-abstraction now (effectively) on $\CalY{\times}(\CalX{\times}\CalX')$ we obtain a hyper in $\Dist^2(\CalX{\times}\CalX')$. (The state-spaces $\CalX$ and $\CalX'$ are the same: we use the prime only to distinguish final and initial.)
\C{We write $\HMMSem{H}$ for the resulting function, and call those $\HMMSem{\cdot}$-images  \emph{abstract \HMM's}.}
\end{definition}
\Def{d1207} extends our earlier definition of \HMM-semantics \cite{McIver:15} by retaining the initial state, so that we can account for collateral effects.%
\footnote{Our earlier domain was $\Dist\CalX{\MFun}\Dist^2\CalX$, as is still the case for channels since do not update their state.}
This new type is the bottom arrow of \Fig{f1041}.

\C{For any abstract \HMM\ $h$} and any collateral type $\CalZ$ we can then define a ``$\CalZ$-lifted'' function $h^{\CalZ}\In\Dist(\CalZ{\times}\CalX){\Fun}\Dist^2(\CalZ{\times}\CalX')$ that takes a collateral correlation $\Pi\In\Dist(\CalZ{\times}\CalX)$ to a hyper in $\Dist^2(\CalZ{\times}\CalX')$ based on such correlations: this is the top arrow of \Fig{f1041}.
\Cf{Need to refer to an appendix section that explains why the arrow at left can go bottom-to-top; or go back to the previous version that uses the conditional with the arrow going down. Which is easier? {\Tx See \App{a1215} for an attempt.} \Cx I have (as you can see) taken a slightly different approach to characterising abstract \HMM's --- unfortunately I cannot yet see the more general approach clearly enough and so am opting for safety. For example, we will probably have to insist that the $h^\CalZ$ is not only sell-defined, but healthy in the earlier sense of \Cite{LiCS15}.}
\C{We then have} 
\begin{definition}\label{d1305}
For \Cr{$h_{1,2}\In\Dist\CalX{\MFun}\Dist^2(\CalX{\times}\CalX')$}{abstract \HMM's $h_{1,2}$} define $h_1;h_2$ to be $h_1\BSemi\,h_2^\CalX$ where (bold) $\BSemi$ is forward Kleisli composition on the (larger) state space $\CalX{\times}\CalX'$, \C{as used for monadic \HMM's} \cite{McIver:15}, and $-^\CalX$ is as in \Fig{f1041} but taking $\CalZ$ to be $\CalX$. (Recall that $\CalX$ and $\CalX'$ are the same.)
\C{It satisfies the essential property that $\HMMSem{H^1;H^2} = \HMMSem{H^1}\BSemi\HMMSem{H^2}$ (\Lem {l1653}).} 
\end{definition}

\begin{figure}
\setlength{\unitlength}{1em}
\definecolor{gray}{rgb}{.5,.5,.5}
\newcommand\Gx {\color{gray}}
\begin{picture}(40,15)(0,-2)
\put(0,0){\makebox(40,15)}
\put(0,0){\makebox(40,12)[t]{\small\parbox[t]{40\unitlength}{
\emph{At left:} Given a collateral correlation $\Pi{\In}\Dist(\CalZ{\times}\CalX)$ let the induced collateral channel $Z\In\CalZ{\MFunR}\CalX$ and $\CalX$-prior $\pi{\In}\Dist\CalX$ be such that $\Pi=Z{\ApplyR}\pi$, that is so that $\Pi_{z,x}=Z_{z,x}\pi_x$.
}
}}
\put(25,0){\makebox(15,9)[t]{\small\parbox[t]{15\unitlength}{
\emph{At right (in grey):} For an initial/final distribution $\delta\In\Dist(\CalX{\times}\CalX')$, the left-multiplication $Z{\MMult}\delta$ produces a distribution in $\Dist(\CalZ{\times}\CalX')$, just as matrix multiplication would.

\smallskip
\emph{At right (in black):} The $\Dist$-lifting (push forward) of the multiplication $(Z\cdot)$ thus takes an initial-final hyper in $\Dist^2(\CalX{\times}\CalX')$ to a hyper in $\Dist^2(\CalZ{\times}\CalX')$.
}
}}

\put(1,-1){\begin{picture}(23,10)
\put(0,0){\makebox(23,10)}
\put(0,0){\makebox(2,2.5)[b]{${\Dist\CalX}$}}
\put(2.6,.7){{\Cx\small abstract \HMM} $h$}
\put(2.5,.4){\vector(1,0){7.5}}
\put(10,0){\makebox(8,2.5)[b]{\raisebox{0pt}[0pt][0pt]{$\Dist(\,\overbrace{\Dist(\CalX{\times}\CalX')}\,)$}}}
\put(1.2,3.5){\makebox(2,3){$Z{\ApplyR}\cdot$}}
\put(1,2.5){\vector(0,1){5}}
\put(16,1.7){\makebox(2,3){\Gx$Z\cdot$}}
\put(15,2.5){\Gx\vector(1,1){2}}
\put(10,3.5){\makebox(2,3){$\Dist(Z\cdot)$}}
\put(13,2.5){\vector(0,1){5}}
\put(0,7.5){\makebox(2,2.5){$\Dist(\CalZ{\times}\CalX)$}}
\put(6.7,9.1){$h^\CalZ$}
\put(4.5,8.8){\vector(1,0){5.5}}
\put(18.5,3.5){\makebox(2,3){\Gx$\Dist(\CalZ{\times}\CalX')$}}
\put(10,7.5){\makebox(8,2.5){$\Dist(\,\Dist(\CalZ{\times}\CalX')\,)$}}
\end{picture}
}
\end{picture}

\smallskip\small

\textbf{\emph{Summary:}} A collateral {\Pf Z} is linked to our state {\Pf X} by joint distribution $\Pi\In\Dist(\CalZ{\times}\CalX)$. This $\Pi$ can be decomposed into its right marginal $\pi{\In}\CalX$ on our state space, and a collateral channel $Z\In\CalZ{\MFunR}\CalX$ between it and $\CalZ$,
{\Cx
i.e\ a right conditional of $\Pi$.
(Right-conditionals are not necessarily unique; but the variation on $x$'s where $\pi.x{=}0$ does not affect $\Dist(Z\cdot)$ at right  (\Lem {l1241}).

Although our abstract $h$ at bottom
can ``see'' only {\Pf X} and prior $\pi$, it can have a collateral effect on {\Pf Z}, given by the ``$\CalZ$-lifting'' $h^\CalZ$ shown at top.
Note that $h^\CalZ$ is not itself an abstract \HMM.
}

\caption{Collateral correlation $\Pi\In\Dist(\CalZ{\times}\CalX)$ lifts $\Dist\CalX\Fun\Dist^2(\CalX{\times}\CalX')$ to $\Dist(\CalZ{\times}\CalX)\Fun\Dist^2(\CalZ{\times}\CalX')$.}
\label{f1041}
\end{figure}

Vulnerabilities $V_g$ are similarly extended, now taking $g\In\CalW{\Fun}(\CalX{\times}\CalX'){\Fun}\Real$, i.e.\ referring to both initial and final states. With them, we define \emph{secure refinement} for \HMM's, that $h_1{\Ref}h_2$ just when $V_g(h_1.\pi)\geq V_g(h_2.\pi)$ for all \C{such $g$'s} and $\pi\In\Dist\CalX$; and \emph{behavioural equivalence} \C{to be} refinement in both directions. We say that (concrete) \HMM's are behaviourally equivalent just when their \C{(abstract)} denotations are. Because refinement $\Ref$ is anti-symmetric \cite{mcivermeinicke10a,Alvim:2014aa}
\C{that establishes \underline{full abstraction} of $\HMMSem{\cdot}$ for that equivalence} on concrete \HMM's.
\C{A basic compositionality result is then that \Def{d1305} respects this equivalence (\Thm{t1329}).}
\Cf{Moved this to here, so that equivalence was defined. (Earlier it was before the previous paraghaph.)}

A second basic compositionality result concerns state extension: an \HMM\ acting on state $\CalX$, i.e.\ on a single variable {\Pf X},  can be also considered to be acting on part of a larger state $\CalZ{\times}\CalX$ with an extra {\Pf Z} (whether correlated or not). Even if the program does not read or write {\Pf Z}, it nevertheless can release information about {\Pf Z}: this is the essence of the collateral ``problem''. The two programs {\Pf leak X{;}\ X:= 0} and simply {\Pf X:= 0} make it clear that the final-state semantics is insufficient for the extension: those two programs both take any $\pi\In\Dist\CalX$ to the same final hyper $[\,[0]\,]$, that {\Pf X} is ``certainly 0''. Yet \C{in the context $\Pf X:= Z;(-)$ they are different}: one reveals {\Pf Z} and the other does not.

\begin{definition}\label{d1345}
Given \C{abstract }$h$ we define its $\CalZ$ extension $h^{{\times}\CalZ}\In\Dist(\CalZ{\times}\CalX)\Fun\Dist^2(\CalZ{\times}\CalX)^2$
in two steps.%
\footnote{We now drop the $\CalX'$ primes, to reduce clutter.}
First, we construct $h^{\CalX{\times}\CalZ^2}$ as in \Fig{f1041}
to produce function a function of type $\Dist(\CalZ^2{\times}\CalX^2)\Fun\Dist^2(\CalZ{\times}\CalX)^2$ that carries the initial value of $\CalX$ as well as two copies of $\CalZ$.
Note that we re-order the set arguments of the $\Dist$'s up to commutativity. Then we define the ``duplicator'' $\zeta\In\CalZ{\times}\CalX\to(\CalZ{\times}\CalX)^2$ as $\zeta.(z,x) = (z,x,z,x)$, and conclude with $h^{\times\CalZ}$ being
$h^{\CalZ^2{\times}\CalX}\circ\Dist\zeta$, which is of type $\Dist(\CalZ{\times}\CalX)\to\Dist^2(\CalZ{\times}\CalX)^2$
as required.
\end{definition}

Our principal result (proved in \App{a1215}) is then
\begin{theorem}\label{t0639C}
Let $h_{1,2}\In\Dist\CalX\to\Dist^2\CalX^2$. Then $h_1\Ref h_2$ iff $\Lift{h_1}{\CalZ}\Ref\Lift{h_2}{\CalZ}$ for every extension $\CalZ$.
\end{theorem}

It is the \emph{only if} of this theorem that shows the extended semantics $\Dist\CalX\to\Dist^2\CalX^2$ is enough to allow $\CalZ$-extension, i.e.\ that our semantics is compositional even in \Cr{Dalenius}{collateral} contexts.

\subsection{Justifying\quad\texttt{REFINEMENT DENIED}\quad due to collateral $g$-vulnerability}\label{s0908}
The definition in \Sec{s1330} of secure-program refinement $\Ref$ looks very strong, mandating a decrease of $g$-vulnerability \emph{for all gain-functions} $g$. Why is it not \emph{too} strong?

A key result of \cite{mcivermeinicke10a} was that $\Ref$ is indeed \underline{not} too strong in closed systems, provided one accepts Bayes vulnerability as a reasonable security measure \cite{Smith:2009aa}, and compositionality as reasonable property of program semantics.%
\footnote{Compositionality is often expressed $\HMMSem{P}{=}\HMMSem{Q}\Implies\HMMSem{{\cal C}(P)}{=}\HMMSem{{\cal C}(Q)}$; but when an (antisymmetric) refinement order is available we refer to the stronger criterion of monotonicity of $\HMMSem{{\cal C}(-)}$ wrt.\ that order.}
The argument was that if some \emph{general}  vulnerability $V_g$, no matter how bizarre, mandates $P{\NRef}Q$ for two programs $P,Q$ (i.e.\ \texttt{\small REFINEMENT DENIED}), then there will be a context $\cal C$ and a prior $\pi$ such that for that $\pi$ the \emph{Bayes} vulnerability of ${\cal C}(P)$, which is not bizarre at all, is strictly less than that of ${\cal C}(Q)$. In \cite{mcivermeinicke10a} the distinguishing $g$ was used
to construct a Markov update $M^g$ acting on $P,Q$'s final state, based on that $g$,  so that $P;M^g$ and $Q;M^g$ were distinguished by the simpler Bayes vulnerability $V_\BVg$.

But our programs \textsf{Lax},\textsf{Strict} from \Fig{f1407} are not distinguished by any $g$ in a closed system: only in an open system, with possible collateral variable {\Pf Z}, do we have $\textsf{Lax}{\NRef}\textsf{Strict}$. That means the technique of \cite{mcivermeinicke10a} will not work here, because no $g$ acting on $\CalX'$, the final state alone, can distinguish the two programs: they produce the same hyper on $\CalX'$. (Recall $\dag$ in \Sec{s1330}.) To make an analogous argument we instead use a collateral-aware $g$ on $\CalX{\times}\CalX'$ to construct a collateral correlation $\Pi^g$ between $\CalZ{=}\CalW$ and $\CalX$; it is then the presence of the induced collateral variable {\Pf W} that allows us to deny the refinement. We can then argue that indeed $\textsf{Lax}$ should not be refined by $\textsf{Strict}$, because in the presence of correlation $\Pi^g$ between {\Pf W} and {\Pf Z} that refinement can be denied on the basis of Bayes vulnerability acting on $\CalW$ alone.

We give more details in \App{a1002}.

\section{Related work; conclusions}
Classical analyses of quantitative information flow assume that the secret does not change, and early approaches to measuring insecurities in programs are based on determining  a ``change in uncertainty'' of some ``prior'' value of the secret --- although how to measure the uncertainty differs in approach.  For example Clark et al \cite{ClarkHM01} use Shannon entropy to estimate the number of bits being leaked; and Clarkson et al \cite{ClarksonMS05}  model a change in belief.  The role of capacity when the prior is not known was stressed by Chatzikokolakis et al \cite{ChatzikokolakisPP08}.   Smith \cite{Smith:2009aa} demonstrated the importance of using measures which have some operational significance, and this idea was developed further \cite{Alvim:2012aa} by introducing the notion of $g$-leakage  to express  leakage wrt.\ very general contexts. The partial order used here on \HMM's is the same  as the  $g$-leakage order introduced by Alvim et al \cite{Alvim:2012aa}, but it  also appeared  in even earlier work \cite{mcivermeinicke10a}.  Its properties have been studied extensively \cite{Alvim:2014aa}.

More recently Marzdiel et al \cite{MardzielAHC14} have analysed information flow of dynamic secrets using a model based on probabilistic automata.  This reflects a view that in computing systems secrets are not necessarily static; our work \cite{mcivermeinicke10a,McIver:15} addresses this idea by providing semantic domains for programs where secrets can change. 

Clark et al \cite{ClarkHM05} give techniques for static analysis of quantitative information flow based on Shannon entropy for a small while-language. Extended \HMM's for modelling side channels have been explored by  Karlof and Wagner \cite{KarlofW03}  and Green et al \cite{GreenNS05} for e.g.\  key recovery.  {\Ax In \Sec{ss1055} our  quantitative capacity bounds on side channels are valid even for collateral leakage. }

The abstract treatment of probabilistic systems with the introduction of a ``refinement order'' was originally due to Jones and Plotkin \cite{Jones:89}; the ideas were extended to include demonic nondeterminism (as well as probability) by us \cite{Morgan:96d}.  In both cases the order (on programs) corresponds to  an order determined by averaging over ``probabilistic predicates'' which are random variables over the state space. The first compositional refinement order for information flow appeared in \cite{mcivermeinicke10a} for security programs expressed in a simple programming language and in \cite{Alvim:2014aa} for a channel model of information flow. 

\noindent{\bf Prospects and future work.}   
In  \cite{McIver:15} we noted that collateral leakage has a significant impact  on abstract-program semantics for quantitative information flow; in this paper we have defined a novel collateral-aware \HMM-style model (\Def{d1305}) that properly takes it into account.
%
The significance of the abstract models in this paper is that even though information flow becomes apparent \emph{only} in a wider context of collateral leakage, our abstract semantics \Def{d1305} and \Thm{t0639C} together show that programs \emph{can still} be compared using only the variables they actually declare in their source code.

 An analogue of \Thm{t0639C} for qualitative information flow appeared in Chen's thesis \cite{Chen:12} where \emph{compositional refinement} checking for  two  programs is reduced to comparisons involving only the declared variables. The benefit of this theorem can be seen in Chen's automation of refinement checking for qualitative information flow. Our general result \Thm{t0639C} holds out the prospect of similar automation for quantitative-security refinement checking.

\newpage
\Tf{References may need to be more uniform (e.g. consistent name abbreviation etc)}
\bibliography{probsNew}


\newpage
\appendix
\section{Summary of terms and notations}\label{s1241}
These entries list in first-use order the points at which notation is introduced during the exposition: a detailed explanation of each is given there.
\Cf{Check entries in the right order.}
{
\newcommand\GI[3] {\>\small#2\>\parbox[t]{27em}{\small#3}\`\small p.\pageref{#1}\\}
\newcommand\Space {~\\[-.5em]}
\begin{tabbing}
\hspace{0em}\=\hspace{9em}\=\kill
\GI{g13928}{collateral leakage}{Leakage from data to which a program does not refer.}
\GI{f1407}{$\CalX$}{(\Fig{f1407}) State space.}
\GI{f1407}{$[{\cdots}]$}{(\Fig{f1407}) Uniform distribution.}
\GI{f1407}{$\From$}{(\Fig{f1407}) Assignment made according to a distribution.}
\GI{f1407}{{\Pf leak}}{(\Fig{f1407}) Emit information from a program that an adversary can observe;
   causes no change in the state of the program.}
\GI{f1407}{${\Pf X}^+,{\Pf X}^-$}{(\Fig{f1407}) Next, previous letter in a list.}
\GI{f1407}{\textsf{Lax},\textsf{Strict}}{(\Fig{f1407}) Example password programs..}
\GI{g135448}{collateral variable}{variable affected by collateral leakage.}
\GI{i1012f}{abstract channel}{Denotation of channel that abstracts from leaked values.}
\GI{i1012f}{abstract \HMM}{Denotation of \HMM\ that abstracts from leaked values.}
\GI{s1540}{$\CalX{\MFun}\CalY$}{Row-stochastic matrix.}
\GI{s1540}{$\CalX{\MFunR}\CalY$}{Column-stochastic (co-stochastic) matrix.}
\GI{s1540}{$C_{x,y},C_{x,-},C_{-,y}$}{Element, row, column of a matrix.}
\GI{g140140}{$\Dist$}{Discrete-distribution- or probability-measure type-constructor.}
\Space
\GI{g1421}{$\pi{\Apply}C$}{Joint distribution on $\CalY{\times}\CalX$ formed by taking prior $\pi$ as input to channel $C$.}
\Space
\GI{g1421}{$[\textrm{joint distribution}]$}{The hyper formed from a joint distribution.}
\GI{g140140}{hyper-distribution}{Type $\Dist^2$; distributions of distributions.}
\GI{g140911}{$\Dist\CalX{\Fun}\Dist^2\CalX$}{Type of abstract channel.}
\GI{s1216}{vulnerability}{Quantitative measure of secret's value.}
\GI{s1216}{gain function}{Used to construct vulnerability functions.}
\GI{s1216}{$\CalW$}{Set from which gain-function's choices are taken.}
\GI{g1416}{$\Exp{\pi}{f}$}{Expected value of function $f$ on distribution $\pi$.}
\GI{g1413}{$g\In\GainF{}\CalX$}{Typical gain function and its type.}
\Space
\GI{g1249}{$\BVg$}{The identity $g$-function, with $\CalW{=}\CalX$, giving $V_\BVg$ as Bayes\\vulnerability.}
\Space
\GI{g1413}{$V_g\In\Dist\CalX{\Fun}\Real$}{Vulnerability $V_g$ induced by a gain function $g$, and its type.}
\GI{g1423}{Bayes Vulnerability}{Greatest probability of guessing a secret on the first try.}
\GI{g1423}{Bayes Risk}{One minus the Bayes Vulnerability.}
\GI{g1428}{$\call_g$ etc.}{Multiplicative $g$-leakage, and various capacities.}
\GI{g1430}{dot $\MMult$}{Matrix multiplication.}
\GI{g1430}{$\leftmarg{\pi},\rightchan{\Pi}$}{left-marginal, right channel of joint distribution}
\GI{s1221}{$\CM{C}{M}$}{\HMM-step, a ``primitive'' \HMM\ with channel $C$ and markov $M$.}
\GI{s1221}{markov (lc.)}{The Markov portion of an \HMM-step.}
\GI{g1434}{semicolon ;}{Sequential composition of \HMM's as matrices.}
\GI{d1429}{$\chan$}{The effective channel of an \HMM.}
\GI{l1029}{$\Par$}{Parallel composition of channels.}
\GI{l1519}{$ \CCap$}{Collateral-capacity estimate.}
\GI{g1519}{$\Uni{\CalX}$}{The uniform distribution over $\CalX$.}
\GI{t1008}{extremal leakage}{An upper bound on leakage.}
\GI{g1426}{$[\textsf{A}\At{\NF{1}{2}},\textsf{B}\At{\NF{1}{4}},\textsf{C}\At{\NF{1}{4}}]$}{Discrete distribution with explicit probabilities.}
\GI{g1251}{wprob}{with probability}
\GI{g1447}{$\Ref$}{Security order on hypers.}
\GI{s1449}{$\HMMSem{\cdot}$}{Semantic function taking \HMM\ matrices to abstract \HMM's.}
\Space
\GI{d1207}{$\pi{\Apply}H$}{Joint distribution on $\CalX{\times}\CalY{\times}\CalX'$ formed by taking prior $\pi$ as input to hyper $H$.}
\Space
\GI{d1207}{$\Dist\CalX{\Fun}\Dist^2\CalX^2$ }{Type of abstract \HMM, sometimes written $\Dist\CalX{\Fun}\Dist^2(\CalX{\times}\CalX')$, with $\CalX{=}\CalX'$, to emphasise the different roles (initial,final) of the two $\CalX$'s.}
\Space
\GI{d1207}{abstract \HMM}{Function in $\Dist\CalX{\Fun}\Dist^2(\CalX{\times}\CalX')$ that is $\HMMSem{H}$ for some $H\In\CalX{\MFun}\CalY{\times}\CalX'$.}
\GI{d1305}{$h^{\CalZ}$}{Utility function for extending collateral-aware \HMM's.}
\GI{d1305}{(bold) semicolon \BSemi}{Sequential composition of abstract \HMM's.}
\GI{d1345}{$h^{{\times}\CalZ}$}{Collateral extension of abstract \HMM\ to collateral variable $\CalZ$.}
\GI{d1345}{$\zeta$}{``Duplicator'' utility function used in the extension $h^{{\times}\CalZ}$.}
\GI{g1255}{$\NC_\CalX$}{Degenerate channel on $\CalX$ that leaks nothing.}
\GI{g1255}{$\Id_\CalX$}{Degenerate Markov transition on $\CalX$ that updates nothing.}
\GI{g1255}{pure channel}{An \HMM\ that (might) leak, but does not update the state.}
\GI{g1255}{pure markov}{An \HMM\ that (might) update the state, but does not leak.}
\GI{g1255}{$\CM{}{M}$}{Abbreviation for $\CM{\NC}{M}$, a pure markov.}
\GI{g1255}{$\CM{C}{}$}{Abbreviation for $\CM{C}{\Id}$, a pure channel.}
\Space
\GI{g1312}{$\Unit$}{The sole element of the unit type $\{\Unit\}$ emitted by a degenerate channel.}
\end{tabbing}
}

\newpage
\section{Detailed calculation of hypers for \Fig{f1407} \AppFrom{from \Sec{s1336}, \Sec{s1330}}}\label{a1105}
\subsection{Elementary models}
For convenience we repeat the programs of \Fig{f1407}, in \Fig{f1407R} here.

Variable {\Pf X} in the program text ranges over the state space ${\CalX} = \{{\Pf A},{\Pf B},{\Pf C}\}$. We begin by modelling its components in the simplest mathematical way, i.e.\ conventionally with probabilistic updates being Markov matrices in $\CalX{\MFun}\CalX$ and leaks being channel matrices in $\CalX{\MFun}\CalY$, where in this simple example $\CalY{=}\CalX$. We then map those different descriptions into the same framework of \HMM-steps, their types becoming $\CalX\Fun\CalY{\times}\CalX$, for varying $\CalY$ (sometimes $\CalX$ itself, and sometimes the unit type $\{\Unit\}$); and finally we use their sequential compositions to get the \HMM\ descriptions of the two programs \textsf{Lax} and \textsf{Strict}.
\begin{figure}
 {\Pf\small
  \begin{tabular}{c}
   // Password X is initially uniformly distributed over ${\CalX} = \{{\Pf A},{\Pf B},{\Pf C}\}$. \\\\[-2ex]
   \begin{tabular}[t]{ll}
    \textsf{``Lax'' user} \\\hline\\[-2.5ex]
    X\From\ [A,B,C] & $^\ast$ \\
    leak [{\Pf X}$^+$,{\Pf X}$^-$] & $^\dagger$ \\
   \end{tabular}
   \hspace{16em}
   \begin{tabular}[t]{ll}
    \textsf{``Strict'' user} \\\hline\\[-2.5ex]
    X\From\ [{\Pf X}$^+$,{\Pf X}$^-$] & $^\$$ \\
    leak [{\Pf X}$^+$,{\Pf X}$^-$] & $^\dagger$ \\
   \end{tabular}\\\\
   \begin{tabular}{l@{~}l}
    $^\ast$    & {\it {\Pf[...]} is the uniform distribution over {\Pf...} ; and {\Pf X\From} assigns to {\Pf X} from a distribution.} \\
    $^\$$      & {\it {\Pf X}$^+$ is the letter following {\Pf X}, and {\Pf X}$^-$ the preceding.} \\[.5ex]
    $^\dagger$ & {\it {\Pf leak [X$^+$,X$^-$]} makes a (fair) choice secretly between {\Pf X}$^+$ or {\Pf  X}$^-$,
                           then emits the value somehow:} \\
               & {\it it does not however indicate whether it chose {\Pf X}$^+$ or {\Pf  X}$^-$.} \\
               & {\it Note that the {\Pf X} referred to in the {\Pf leak} statement is the updated value, after the {\Pf X\From}~.}
   \end{tabular}
  \end{tabular}
 }

\bigskip
{\small \textsf{Lax} may choose any new password, uniformly, including his current; but \textsf{Strict} must \emph{change} his password, again uniformly. In both cases the distribution of the new {\Pf X} is again uniform: for \textsf{Lax} it is independent of {\Pf X}'s initial value; but for \textsf{Strict} is is correlated.

Both users, in the second statement, suffer an ``over the shoulder'' attack against the new password.}
\caption{Updating a password (repeat of \Fig{f1407})}\label{f1407R}
\end{figure}

The first statement {\Pf X\From\ [A,B,C]} of \textsf{Lax} corresponds to a Markov matrix in $\CalX{\MFun}\CalX$ with the input at left and output at top:
\[
 M^{L1}:\hspace{5em}
 \STRUT\left(
  \begin{array}{l@{}ccc}
   \ROW{\Pf A}&\COL{\Pf A}\NF{1}{3} & \COL{\Pf B}\NF{1}{3} & \COL{\Pf C}\NF{1}{3} \\
   \ROW{\Pf B}&\NF{1}{3} & \NF{1}{3} & \NF{1}{3} \\
   \ROW{\Pf C}&\NF{1}{3} & \NF{1}{3} & \NF{1}{3} \\
  \end{array}
 \right)
 \hspace{3em}\parbox{10em}{For \textsf{Lax} the input is ignored; the output is a uniform choice over $\CalX$.}
\]
However the first statement {\Pf X\From\ [{\Pf X}$^+$,{\Pf X}$^-$]} of \textsf{Strict} corresponds this Markov matrix instead:
\[
M^{S1}:\hspace{5em}
\STRUT\left(
  \begin{array}{l@{}ccc}
   \ROW{\Pf A}& \COL{\Pf A}0 & \COL{\Pf B}\NF{1}{2} & \COL{\Pf C}\NF{1}{2} \\
   \ROW{\Pf B}& \NF{1}{2} & 0 & \NF{1}{2} \\
   \ROW{\Pf C}& \NF{1}{2} & \NF{1}{2} & 0 \\
  \end{array}
 \right)
 \hspace{3em}\parbox{10em}{For \textsf{Strict} the output is a uniform choice over anything \emph{but} the input.}
\]
The second statement {\Pf X\From\ [A,B,C]} in both cases corresponds to a channel matrix in $\CalX{\MFun}\CalY$, where in fact $\CalY{=}\CalX$ because the observables are of the same type as the state:
\[
 C^{2}:\hspace{5em}
 \STRUT\left(
  \begin{array}{l@{}ccc}
   \ROW{\Pf A}& \COL{\Pf A}0 & \COL{\Pf B}\NF{1}{2} & \COL{\Pf C}\NF{1}{2} \\
   \ROW{\Pf B}& \NF{1}{2} & 0 & \NF{1}{2} \\
   \ROW{\Pf C}& \NF{1}{2} & \NF{1}{2} & 0 \\
  \end{array}
 \right)
 \hspace{3em}\parbox{12em}{The over-the-shoulder leak is uniformly any value not equal to the current state.}
\]

\subsection{Elementary models make \HMM\ steps}
To build the two programs from their components, we must convert those components to \HMM\ steps: note that although the components are all matrices, Markov matrices and channels are converted in different ways. We start with the conversion of the markovs: converted from $\CalX{\MFun}\CalX$ to the \HMM-type $\CalX{\MFun}\CalY{\times}\CalX$ they become
\[
 \STRUT\left(
  \begin{array}{l@{}ccc}
   \ROW{\Pf A}&\COL{\Unit\Pf A}\NF{1}{3} & \COL{\Unit\Pf B}\NF{1}{3} & \COL{\Unit\Pf C}\NF{1}{3} \\
   \ROW{\Pf B}&\NF{1}{3} & \NF{1}{3} & \NF{1}{3} \\
   \ROW{\Pf C}&\NF{1}{3} & \NF{1}{3} & \NF{1}{3} \\
  \end{array}
 \right)
\hspace{8em}
 \STRUT\left(
  \begin{array}{l@{}ccc}
   \ROW{\Pf A}&\COL{\Unit\Pf A}0 & \COL{\Unit\Pf B}\NF{1}{2} & \COL{\Unit\Pf C}\NF{1}{2} \\
   \ROW{\Pf B}&\NF{1}{2} & 0 & \NF{1}{2} \\
   \ROW{\Pf C}&\NF{1}{2} & \NF{1}{2} & 0 \\
  \end{array}
 \right)
\]
where e.g.\ the label \Unit{\Pf A} means the $y,x'$ pair $(\Unit,{\Pf A})$ --- the observable is ``unit'', the single element of the unit type $\{\Unit\}$,
\label{g1312}
and the new state is {\Pf A} because, these markovs' being leak-free, they (in effect) have no observables (equivalently, one always observes the same thing, the \Unit, so no information flows).
\footnote{In the notation of \App{a1005} they would be $\CM{}{M^{L1}}$ and  $\CM{}{M^{S1}}$, both of \HMM-type $\CalX{\MFun}\{\Unit\}{\times}\CalX$~.}

Conversion of the channel(s) does however leak information, giving in both cases
\footnote{In the notation of \App{a1005} this would be $\CM{C^2}{}$ of \HMM-type $\CalX{\MFun}\CalY{\times}\CalX$ where, recall in fact $\CalY{=}\CalX$}
\[
 \STRUT\left(
  \begin{array}{l@{}ccccccccc}
   & \COL{\Pf AA}~~~ & \COL{\Pf AB}~~~ & \COL{\Pf AC}~~~ & \COL{\Pf BA}~~~ & \COL{\Pf BB}~~~ & \COL{\Pf BC}~~~ & \COL{\Pf CA}~~~ & \COL{\Pf CB}~~~ & \COL{\Pf CC}~~~ \\[-1em]
   \ROW{\Pf A}& 0 & 0 & 0 & \NF{1}{2} & 0 & 0 & \NF{1}{2} & 0 & 0 \\
   \ROW{\Pf B}& 0 & \NF{1}{2} & 0 & 0 & 0 & 0 & 0 & \NF{1}{2} & 0 \\
   \ROW{\Pf C}& 0 & 0 & \NF{1}{2} & 0 & 0 & \NF{1}{2} & 0 & 0 & 0 \\
  \end{array}
 \right)
\hspace{2em}\parbox{17em}{\small The columns are labelled with pairs $(y,x')$, so that e.g.\ from input {\Pf A} there are two possible results: leak {\Pf B} and stay in state {\Pf A}, or leak {\Pf C} and stay in state {\Pf A}.}
\]

\subsection{Sequential composition of \HMM-steps}
To get the \HMM\ representation of the two programs, we use the definition of \HMM\ sequential composition \Eqn{e1555} in \Sec{s1735}, which we repeat here for convenience:
\begin{equation}\label{e1734}
 (H^1;H^2)_{x,(y^1,y^2),x'}
 \Wide{=}
 \sum_{x''} H^1_{x,y^1,x''}H^2_{x'',y^2,x'} ~.
 \hspace{10ex}\textrm{recalling \Eqn{e1555}}
\end{equation}
We have two versions of $H^1$, one for each of \textsf{Lax} and \textsf{Strict}, and a single version of $H^2$ that applies to both. Type $\CalY^1$ is the unit $\{\Unit\}$; and type $\CalY^2$ is the same as $\CalX$. Our composite \HMM\ (no longer simply an \HMM-step) will have observables' type $\CalY{=}\CalX$, because we can drop the unit type;
\footnote{The Cartesian product $\{\Unit\}{\times}\CalY$ is isomorphic to $\CalY$ alone: we have \HMM-composed $\CalX{\MFun}\{\Unit\}{\times}\CalX$ and $\CalX{\MFun}\CalY{\times}\CalX$ to get $\CalX{\MFun}(\{\Unit\}{\times}\CalY){\times}\CalX$, isomorphically $\CalX{\MFun}\CalY{\times}\CalX$.}
and its output type is still $\CalX$. For \textsf{Lax} the result is this \HMM, obtained by applying \Eqn{e1555} to \textsf{Lax}'s $H^1$ and the common $H^2$:
\[
 \STRUT\left(
  \begin{array}{l@{}ccccccccc}
   & \COL{\Pf AA}~~~ & \COL{\Pf AB}~~~ & \COL{\Pf AC}~~~ & \COL{\Pf BA}~~~ & \COL{\Pf BB}~~~ & \COL{\Pf BC}~~~ & \COL{\Pf CA}~~~ & \COL{\Pf CB}~~~ & \COL{\Pf CC}~~~ \\[-1em]
   \ROW{\Pf A}& 0 & \NF{1}{6} & \NF{1}{6} & \NF{1}{6} & 0 & \NF{1}{6} & \NF{1}{6} & \NF{1}{6} & 0 \\
   \ROW{\Pf B}& 0 & \NF{1}{6} & \NF{1}{6} & \NF{1}{6} & 0 & \NF{1}{6} & \NF{1}{6} & \NF{1}{6} & 0 \\
   \ROW{\Pf C}& 0 & \NF{1}{6} & \NF{1}{6} & \NF{1}{6} & 0 & \NF{1}{6} & \NF{1}{6} & \NF{1}{6} & 0 \\
  \end{array}
 \right)
 \hspace{2em}\parbox{17em}{\small For \textsf{Lax} the rows are identical, because the input is ignored; the result is uniformly distributed over outcomes in which the leaked value and the new state are different.}
\]
For \textsf{Strict} the result is this \HMM, this time using \textsf{Strict}'s $H^1$ (and again the common $H^2$):
\[
 \STRUT\left(
  \begin{array}{l@{}ccccccccc}
     & \COL{\Pf AA}~~~ & \COL{\Pf AB}~~~ & \COL{\Pf AC}~~~ & \COL{\Pf BA}~~~ & \COL{\Pf BB}~~~ & \COL{\Pf BC}~~~ & \COL{\Pf CA}~~~ & \COL{\Pf CB}~~~ & \COL{\Pf CC}~~~ \\[-1em]
   \ROW{\Pf A}& 0 & \NF{1}{4} & \NF{1}{4} & 0 & 0 & \NF{1}{4} & 0 & \NF{1}{4} & 0 \\
   \ROW{\Pf B}& 0 & 0 & \NF{1}{4} & \NF{1}{4} & 0 & \NF{1}{4} & \NF{1}{4} & 0 & 0 \\
   \ROW{\Pf C}& 0 & \NF{1}{4} & 0 & \NF{1}{4} & 0 & 0 & \NF{1}{4} & \NF{1}{4} & 0 \\
  \end{array}
 \right)
 \hspace{2em}\parbox{17em}{\small For \textsf{Strict} the rows are \emph{not} identical, because the first \HMM-step updates the state in a way dependent on its incoming value.}
\]

\subsection{The semantic view: hypers for \textsf{Lax}}
We are now interested in the hypers produced by each of the two \HMM's just above when they are applied to the uniform prior $[{\Pf A},{\Pf B},{\Pf C}]$ that assigns probability $\NF{1}{3}$ to each of {\Pf A},{\Pf B},{\Pf C}, an assumption we have made for this example). We do \textsf{Lax} first: the joint distribution in $\Dist(\CalX{\times}\CalY{\times}\CalX')$, where we are now referring to the final state with a prime, is obtained by multiplying each row of the \HMM\ by the corresponding prior probability (in this case $\NF{1}{3}$ for all of them), giving
\[
 \STRUT\left(
  \begin{array}{l@{}ccccccccc}
   & \COL{\Pf AA}~~~ & \COL{\Pf AB}~~~ & \COL{\Pf AC}~~~ & \COL{\Pf BA}~~~ & \COL{\Pf BB}~~~ & \COL{\Pf BC}~~~ & \COL{\Pf CA}~~~ & \COL{\Pf CB}~~~ & \COL{\Pf CC}~~~ \\[-1em]
   \ROW{\Pf A}& 0 & \NF{1}{18} & \NF{1}{18} & \NF{1}{18} & 0 & \NF{1}{18} & \NF{1}{18} & \NF{1}{18} & 0 \\
   \ROW{\Pf B}& 0 & \NF{1}{18} & \NF{1}{18} & \NF{1}{18} & 0 & \NF{1}{18} & \NF{1}{18} & \NF{1}{18} & 0 \\
   \ROW{\Pf C}& 0 & \NF{1}{18} & \NF{1}{18} & \NF{1}{18} & 0 & \NF{1}{18} & \NF{1}{18} & \NF{1}{18} & 0 \\
  \end{array}
 \right)
\hspace{2em}\parbox{15em}{\small To convert this to a hyper on $\CalX{\times}\CalX'$, we abstract from $\CalY$, the first component of each column.}
\]

Because of the symmetry, to abstract from $\CalY$ we can concentrate on just one observed value, say {\Pf A}. In that case, the posterior in $\Dist(\CalX{\times}\CalX')$ is $[\,[\,{\Pf AB},{\Pf AC}\,],[\,{\Pf BB},{\Pf BC}\,],[\,{\Pf CB},{\Pf CC}\,]\,]$, where the first component is the row label $x$ from just above, and the second component is the second component of the column label $x'$ --- the observation $y{=}{\Pf A}$ has been abstracted. We can write that more concisely in this case as the product $[\,{\Pf A},{\Pf B},{\Pf C}\,]{\times}[\,{\Pf B},{\Pf C}\,]$ of two (uniform) distributions, one in $\Dist\CalX$ and the other in $\Dist\CalX'$. 

For observation {\Pf B}, analogously we get  $[\,{\Pf A},{\Pf C}\,]{\times}[\,{\Pf A},{\Pf C}\,]$; for observation {\Pf C} we get  $[\,{\Pf A},{\Pf C}\,]{\times}[\,{\Pf A},{\Pf B}\,]$. Since the observations (in this case) have equal (marginal) probabilities, the overall hyper in $\Dist^2(\CalX{\times}\CalX')$ is uniform: it is
\begin{equation}\label{e1743}
 [\,[\,{\Pf A},{\Pf B},{\Pf C}\,]{\times}[\,{\Pf B},{\Pf C}\,],\quad[\,{\Pf A},{\Pf B},{\Pf C}\,]{\times}[\,{\Pf A},{\Pf C}\,],\quad[\,{\Pf A},{\Pf B},{\Pf C}\,]{\times}[\,{\Pf A},{\Pf B}\,]\,]~.
\end{equation}
If we project \Eqn{e1743} onto its second component we get a hyper in $\Dist^2\CalX'$ on the final state alone: it is
\(
 [\,[\,{\Pf B},{\Pf C}\,],\,[\,{\Pf A},{\Pf C}\,],\,[\,{\Pf A},{\Pf B}\,]\,]~,
\)
so that for \textsf{Lax} we will wprob $\NF{1}{3}$ believe a-posteriori that $x'$ is uniformly either {\Pf B} or {\Pf C}, when we observed {\Pf A}; for observations {\Pf B},{\Pf C} resp.\ our posteriors in $\Dist\CalX'$ alone are $[\,{\Pf A},{\Pf C}\,]$ and $[\,{\Pf A},{\Pf B}\,]$ resp.

If however we project onto the first component, we get posterior $[\,{\Pf A},{\Pf B},{\Pf C}\,]$ in all three cases: we have learned nothing about the \emph{initial} value of $x$.

\subsection{The semantic view: hypers for \textsf{Strict}}
For \textsf{Strict}, sensitive to the initial state, the results are different. First, the joint distribution in $\Dist(\CalX{\times}\CalY{\times}\CalX')$ is again obtained by multiplying the uniform prior through, giving now
\[
 \STRUT\left(
  \begin{array}{l@{}ccccccccc}
   & \COL{\Pf AA}~~~ & \COL{\Pf AB}~~~ & \COL{\Pf AC}~~~ & \COL{\Pf BA}~~~ & \COL{\Pf BB}~~~ & \COL{\Pf BC}~~~ & \COL{\Pf CA}~~~ & \COL{\Pf CB}~~~ & \COL{\Pf CC}~~~ \\[-1em]
   \ROW{\Pf A}& 0 & \NF{1}{12} & \NF{1}{12} & 0 & 0 & \NF{1}{12} & 0 & \NF{1}{12} & 0 \\
   \ROW{\Pf B}& 0 & 0 & \NF{1}{12} & \NF{1}{12} & 0 & \NF{1}{12} & \NF{1}{12} & 0 & 0 \\
   \ROW{\Pf C}& 0 & \NF{1}{12} & 0 & \NF{1}{12} & 0 & 0 & \NF{1}{12} & \NF{1}{12} & 0 \\
  \end{array}
 \right)
\]
Again we concentrate on just one observation {\Pf A}, in which case, the posterior in $\Dist(\CalX{\times}\CalX')$ is $[\,{\Pf AB},{\Pf AC},{\Pf BC},{\Pf CB}\,]$.%
\footnote{Take each of the non-zero entries under a column with left component {\Pf A}; retain only $x$ and $x'$, dropping the {\Pf A}; and, since all the probabilities are the same, form the uniform distribution on that.}
For observation {\Pf B}, analogously we get  $[\,{\Pf AC},{\Pf BA},{\Pf BC},{\Pf CA}\,]$; for observation {\Pf C} we get  $[\,{\Pf AB},{\Pf BA},{\Pf CA},{\Pf CB}\,]$. Since the observations (again) have equal  probabilities, the overall hyper in $\Dist^2(\CalX{\times}\CalX')$ is again uniform:
\begin{equation}\label{e1719}
 [\,[\,{\Pf AB},{\Pf AC},{\Pf BC},{\Pf CB}\,],\quad[\,{\Pf AC},{\Pf BA},{\Pf BC},{\Pf CA}\,],\quad[\,{\Pf AB},{\Pf BA},{\Pf CA},{\Pf CB}\,]\,]~.
\end{equation}

Now if we project this hyper, for \textsf{Strict}, onto its second component (the final state), we get the same hyper in $\Dist^2\CalX'$ as for \textsf{Lax} just above. But if we project onto the first component (the initial state), this time we get posterior 
\begin{equation}\label{e1721}
 [\,[\,{\Pf A},{\Pf A},{\Pf B},{\Pf C}\,],\quad[\,{\Pf A},{\Pf B},{\Pf B},{\Pf C}\,],\quad[\,{\Pf A},{\Pf B},{\Pf C},{\Pf C}\,]\,]~,
\end{equation}
which we constructed just by erasing the second component of each pair in \Eqn{e1719}. The ``uniform distribution'' with repeated components in the list gives proportionally more probability to the repetitions: written in the conventional way, for \Eqn{e1721} we would get
\begin{equation}\label{e1722}
 [\,[\,{\Pf A}\At{\NF{1}{2}},{\Pf B}\At{\NF{1}{4}},{\Pf C}\At{\NF{1}{4}}\,],\quad[\,{\Pf A}\At{\NF{1}{4}},{\Pf B}\At{\NF{1}{2}},{\Pf C}\At{\NF{1}{4}}\,],\quad[\,{\Pf A}\At{\NF{1}{4}},{\Pf B}\At{\NF{1}{4}},{\Pf C}\At{\NF{1}{2}}\,]\,]~,
\end{equation}
which is indeed the same as \Eqn{e0906} in \Sec{s1330} above.

\subsection{The semantics of composition, and relative security}
The above constructions started from Markov- and channel matrices, converted them into \HMM-step matrices, and then sequentially composed those matrices according to \Eqn{e1734}. The semantic alternative is to take the denotations at the very beginning, i.e.\ the abstract \HMM's corresponding to the individual program fragments, and then to compose them according to the \emph{semantic} definition at \Def{d1305}. The essential property $\HMMSem{H^1;H^2} = \HMMSem{H^1};\HMMSem{H^2}$ referred to there ensures that the outcome would have been the same. Thus, either way, we can summarise all the above by saying that \textsf{Lax}, as an abstract \HMM, takes prior $[\,{\Pf A},{\Pf B},{\Pf C}\,]$ to hyper \Eqn{e1743} which, when projected onto the initial component $\CalX$ is $[\,[\,{\Pf A},{\Pf B},{\Pf C}\,]\,]$ --- showing that indeed \textsf{Lax} releases nothing about the initial state. The same procedure with \textsf{Strict} however gives the initial-state hyper \Eqn{e1722} which is \emph{strictly less secure} than $[\,[\,{\Pf A},{\Pf B},{\Pf C}\,]\,]$ in the hyper-order $\Ref$. This can be demonstrated even with Bayes Vulnerability $V_{g_{\mathit{id}}}$, which is $\NF{1}{3}$ for $[\,[\,{\Pf A},{\Pf B},{\Pf C}\,]\,]$ but $\NF{1}{2}$ for \Eqn{e1722} --- but, more significantly, that strict-less-secure relation means that \emph{for any $g$} applying $V_g$ to $[\,[\,{\Pf A},{\Pf B},{\Pf C}\,]\,]$ will give no worse (i.e.\ no higher) vulnerability than applying it to \Eqn{e1722} or, equivalently, to \Eqn{e0906}.

\newpage
\section{Proofs for \Sec{s1221}}\label{a1626}
\subsection*{Proof for \Lem{l1029} \AppFrom{from \Sec{s1101}}}
\Cf{Why is this material all in Annabelle's colour?}
{\Ax
Let $H$ be an \HMM\ and $\CM{C}{M}$ an \HMM-step. Then 
\[
 \begin{array}{lllp{10em}}
  \chan.\,\CM{C}{M}   &=& C \\
  \chan.\,(\CM{C}{M};H) &=& C\Par(M{\MMult}\chan.H)
\end{array}
\]
where \C{in general $(C^1{\Par}C^2)_{x,(y^1,y^2)} = C^1_{x,y^1}C^2_{x,y^2}$} is parallel composition of channels. The $M$ cannot be discarded, since it affects the prior of the ``tail'' $H$ of the sequential composition.
\begin{proof}
We prove each equality independently:

\begin{Reason}
	\Step{}{
		(\chan.\CM{C}{M})_{x,y}
	}
	\StepR{$=$}{Def.~$\chan$}{
		\sum_{x'} \CM{C}{M}_{x,y,x'}
	}
	\StepR{$=$}{Def.~\HMM-matrix}{
		\sum_{x'} C_{x,y}M_{x,x'}
	}
	\StepR{$=$}{$M$ is stochastic}{
		C_{x,y}
	}
\end{Reason}

\begin{Reason}
	\Step{}{
		(\chan.(\CM{C}{M};H))_{x,(y_1, \dots , y_{n})}
	}
	\StepR{$=$}{Def.~$\chan$}{
		\sum_{x'} (\CM{C}{M};H)_{x,(y_1, \dots , y_{n}), x'}
	}
	\StepR{$=$}{Def.~$(;)$}{
		\sum_{x'} \sum_{x''} \CM{C}{M}_{x,y_1, x''} H_{x'', (y_2, \dots , y_{n}), x'}
	}
	\StepR{$=$}{Def.~\HMM-matrix}{
		\sum_{x'} \sum_{x''} C_{x,y_1} M_{x, x''} H_{x'', (y_2, \dots , y_{n}), x'}
	}
	\StepR{$=$}{Move stuff around}{
		C_{x,y_1} \sum_{x''}  M_{x, x''} \sum_{x'} H_{x'', (y_2, \dots , y_{n}), x'}
	}
	\StepR{$=$}{Def.~$\chan$}{
		C_{x,y_1} \sum_{x''}  M_{x, x''} (\chan.H)_{x'', (y_2, \dots , y_{n})}
	}
	\StepR{$=$}{Matrix multiplication}{
		C_{x,y_1} (M\CProd \chan.H)_{x, (y_2, \dots , y_{n})}
	}
	\StepR{$=$}{Parallel composition}{
		(C;(M\CProd \chan.H))_{x, (y_1, y_2, \dots , y_{n})}\quad.
	}
\end{Reason}

\end{proof}
}

\subsection*{Proof for \Lem{l1519} \AppFrom{from \Sec{s1101}}}

{\Ax
For any $H$ let $\CCap.H$ be defined
\[
 \begin{array}{lll@{\hspace{4.8em}}r}
 \CCap.\,\CM{C}{M}     &=& \call_\forall(\forall,C)
 & \textrm{if $H{=}\CM{C}{M}$} \\
  \CCap.\,(\CM{C}{M};H) &=& \call_\forall(\forall,C) + \min(\call_\forall(\forall, M), \CCap.H)~.
    & \textrm{if $H{=}\CM{C}{M};H'$}
\end{array}
\]
Then $\call_\forall(\forall, \chan.H) \le \CCap.H$ where we interpret the stochastic matrix $M$ (rhs)
as a channel.
\begin{proof}

We use induction in the number of \HMM-steps of $H$.

Base case: $H$ has only one step.
\begin{Reason}
	\Step{}{
		\call_\forall(\forall, \chan.\CM{C}{M})
	}
	\StepR{$=$}{Def.~$\chan$}{
		\call_\forall(\forall, C)
	}
	\StepR{$=$}{Def.~$\CCap$}{
		\CCap.\,\CM{C}{M}
	\quad .
	}
\end{Reason}

Inductive step: assume the lemma is true for \HMM's composed of $n$ steps. Let $H = \CM{C}{M};H'$, where $H'$ has $n$ steps.
\begin{Reason}
	\Step{}{
		\call_\forall(\forall, \chan.(\CM{C}{M};H'))
	}
	\StepR{$=$}{Def.~$\chan$}{
		\call_\forall(\forall, C \parallel (M \CProd \chan.H'))
	}
	\StepR{$=$}{\cite[Thm.~5.1]{Alvim:2012aa}}{
		\call_\BVg(\Uni{\CalX}, C \parallel (M \CProd \chan.H'))
	}
	\StepR{$\le$}{\cite[Cor.~7]{Espinoza:2013aa}}{
		\call_\BVg(\Uni{\CalX}, C) +  \call_\BVg(\Uni{\CalX}, M \CProd \chan.H')
	}
	\StepR{$\le$}{\cite[Thm.~6]{Espinoza:2013aa}}{
		\call_\BVg(\Uni{\CalX}, C) +  \min(\call_\BVg(\Uni{\CalX}, M),  \call_\BVg(\Uni{\CalX}, \chan.H'))
	}
	\StepR{$=$}{\cite[Thm.~5.1]{Alvim:2012aa}}{
		\call_\BVg(\Uni{\CalX}, C) +  \min(\call_\BVg(\Uni{\CalX}, M),  \call_\forall(\forall, \chan.H'))
	}
	\StepR{$\le$}{Inductive hypothesis}{
		\call_\BVg(\Uni{\CalX}, C) +  \min(\call_\BVg(\Uni{\CalX}, M),  \CCap.H')
	}
		\StepR{$\le$}{monotonicity}{
		\call_\forall(\forall, C) +  \min(\call_\forall(\forall, M),  \CCap.H')
	}
	\StepR{$=$}{Def.~$\CCap$}{
		\CCap.\,(\CM{C}{M};H')
		\quad.
	}
\end{Reason}
\Af{Added a monotonicity step.}

\end{proof}

}

\subsection*{Example showing that \Lem{l1519} can be conservative  \AppFrom{from \Sec{s1101}}}

Consider the following program fragment. 

{\tt\small
\begin{tabbing}
 // \textit{Prog: \XS\ is \C{initialised} uniformly at random.}\\
 xs:= xs $\PC{{\nicefrac{1}{3}}}$ $\neg$xs~; \\
 leak xs[0] $\PC{{\nicefrac{1}{2}}}$ xs[1]~;\\
 xs:= xs $\PC{{\nicefrac{1}{2}}}$ $\neg$xs\\
\end{tabbing}}

Hidden variable {\tt xs} is set initially uniformly at random over the four possible arrays of length two having $0,1$ entries.
Next the array values are all flipped (probability $2/3$) or all left alone (probability $1/3$); afterwards the value of either the first or second bit is leaked, but only the value is observed. Finally {\tt xs} is updated again. 

We model this as \Cr{two}{three} \HMM\ basic steps : \C{$H \Defs \CM{}{M_1}; \CM{C_2}{}; \CM{}{M_2}$}, where $M_1$ corresponds to the first update of {\tt xs}, $C_2$ corresponds to the {\tt leak} statement and $M_2$ the last update of {\tt xs}. \C{But from \App{a1005} we can combine the second and third, writing it as just the two steps $\CM{}{M_1}; \CM{C_2}{M_2}$.} Applying \Lem{l1519} we find

\begin{Reason}
\Step{}
{ \CCap.\,H}
\Step{$=$}
{\call_\BVg(\Uni{\CalX},I) + \min(\call_\BVg(\Uni{\CalX}, M_1), \CCap.(\CM{C_2}{M_2}))}
\StepR{$=$}{Identity channel leaks nothing; \Lem{l1519}}
{ \min(\call_\BVg(\Uni{\CalX}, M_1), \call_\BVg(\Uni{\CalX}, C_2))~.}
\end{Reason}

We compute the values of the relevant matrices as follows:

\[
M_1 \Defs \left(\begin{array}{cccc}
    1/3 &  0 &   0 &  2/3\\
    0  & 1/3 & 2/3&  0\\
    0 &  2/3  & 1/3 & 0 \\
    2/3  &  0 &   0 & 1/3
\end{array}\right)
\quad\quad
C_2 \Defs \left(\begin{array}{cc}
    1 &  0 \\
    1/2  & 1/2\\
    1/2 &  1/2 \\
    0  &  1 
\end{array}\right)
\]

Finally we compute the leakages:
\[
\call_\BVg(\Uni{\CalX}, M_1) \Wide{=} \lg( (1/6 + 1/6 + 1/6 + 1/6)*4) \Wide{=} \lg(8/3) \Wide{=} 1.415~,
\]
\[
\call_\BVg(\Uni{\CalX}, C_2) \Wide{=} \lg( (1/4 + 1/4)*4) \Wide{=} \lg(2) \Wide{=} 1 ~.
\]
Hence $\CCap.\,H = 1$.

However we can get an exact computation of leakage by calculating $\chan.H$ exactly via \Lem{l1029}, which gives
$\call_\forall(\forall, \chan.H) = \call_\BVg(\Uni{\CalX}, M_1\cdot C_2)$ where

\[
M_1\cdot C_2 \Defs \left(\begin{array}{cc}
    1/3 &  2/3 \\
    1/2  & 1/2\\
    1/2 &  1/2 \\
    2/3  &  1/3 
\end{array}\right)
\]

So finally we have $\call_\forall(\forall, \chan.H) = \lg((1/6 + 1/6)4) = \lg(4/3) = 0.415$, which is considerably less than $\CCap.\,H$.

\subsection*{Proof for \Thm{t1008} \AppFrom{from \Sec{s1101}}}

Given $H$ and $\Pi\In\Dist(\CalZ{\times}\CalX)$ with $\leftmarg\pi,\rightmarg\pi$ resp.\  the marginals 
of \/ $\Pi$ on $\CalZ,\CalX$, define conditional $\rightchan\Pi$ as in \Sec{s1124}, there exists $\hat{g}\In\GainF{\CalZ}\CalZ$ and $\hat{g}^{\joint}\In \GainF{\CalZ}\CalX$ such that
\[
\call_{\forall}(\leftmarg\pi, \rightchan\Pi{\MMult} \chan.H) \Wide{=} \call_{\hat{g}}(\leftmarg\pi, \rightchan\Pi{\MMult} \chan.H) \Wide{=} \call_{\hat{g}^{\joint}} (\rightmarg\pi, \chan.H)~.
\]

\begin{proof}
Let $a \Defs \min_{z\In\CalX, \leftmarg\pi_z>0}\leftmarg\pi_z$, that is, $a$ is the minimum non-zero probability in $\leftmarg\pi$.
The first equality follows from \cite{Alvim:2014aa}[Theorem 10], with  
$
\hat{g}.z'.z = {a}/{\margz_z}
$, whenever $z'=z$ and $\margz_z>0$, and is zero otherwise.

We prove the second equality as follows. Let $C \Defs \chan.H$ and let $\chanx$ in $\CalX{\MFun}\CalZ$ be such that $\Pi_{z,x} = \rightmarg\pi_x\leftchan\Pi_{x,z}$.
Next, given $g \In \GainF{\cal W}\CalZ$, define $g^{\joint}\In  \GainF{\cal W}\CalX$ 
by
$
g^\joint .w.x = \sum_{z\In\CalZ} \chanx_{x,z} g.w.z~.
$

We now reason as follows  that $V_{g}[\margz{\Apply}\chanz{\CProd} C] = V_{g^\joint}[\margx{\Apply}C]$.
\begin{Reason}
\Step{}{
V_{g}[\margz{\Apply}\chanz{\CProd} C]
}
\StepR{$=$}{Def.~$V_g$}{
\sum_{y\In\CalY} \max_{w\In\calw} \sum_{z\In\CalZ} \margz_z (\chanz\CProd C)_{z,y}  g.w.z
}
\StepR{$=$}{cascading}{
\sum_{y\In\CalY} \max_{w\In\calw} \sum_{z\In\CalZ} \margz_z (\sum_{x\In\CalX}\chanz_{z,x} C_{x,y}) g.w.z
}
\StepR{$=$}{move stuff}{
\sum_{y\In\CalY} \max_{w\In\calw} \sum_{z\In\CalZ, x\In\CalX} (\margz_z \chanz_{z,x})C_{x,y}  g.w.z
}
\StepR{$=$}{$\margz_z \chanz_{z,x} = \joint_{z,x} = \margx_x \chanx_{x,z}$}{
\sum_{y\In\CalY} \max_{w\In\calw} \sum_{z\In\CalZ, x\In\CalX} (\margx_x \chanx_{x,z})C_{x,y}  g.w.z
}
\StepR{$=$}{move stuff}{
\sum_{y\In\CalY} \max_{w\In\calw} \sum_{x\In\CalX} \margx_x C_{x,y} (\sum_{z\In\CalZ} \chanx_{x,z}  g.w.z)
}
\StepR{$=$}{Def.~$g^\joint$}{
\sum_{y\In\CalY} \max_{w\In\calw} \sum_{x\In\CalX} \margx_x C_{x,y}  g^\joint.w.x
}
\StepR{$=$}{Def.~$V_{g^\joint}$}{
V_{g^\joint}[\margx{\Apply}C]~.
}\
\end{Reason}

The proof of $V_{g}[\margz] = V_{g^\joint}[\margx]$ is similar, implying that $\call_{\hat{g}}(\margz, \chanz\CProd\chan.H) = \call_{\hat{g}^{\joint}} (\margx, \chan.H)$ as required.
\end{proof}

\subsection*{Proof for \Thm{t1203} \AppFrom{from \Sec{s1101}}}

Given $H$ and $\Pi$ as above,
we have
\[
\call_\forall(\leftmarg\pi, \chanz{\MMult}\chan.H)
\Wide{\le}
\call_\BVg(\Uni{\CalX}, \chan.H)~,
\]
where $\leftmarg\pi,\rightchan\Pi$ are as defined in \Thm{t1008},
and $\Uni{\CalX}$ is the uniform probability distribution over $\CalX$.

\begin{proof}
It is enough to show the following:
\[
\call_\forall(\leftmarg\pi, \chanz{\MMult}\chan.H)
\Wide{\le}
\call_\forall(\rightmarg\pi, \chan.H) \Wide{=} 
\call_\BVg(\Uni{\CalX}, \chan.H)
\]
The first inequality follows from the right capacity bound from \cite[Cor 23]{Alvim:2014aa}; the equality then follows from  \cite[Thm 10]{Alvim:2014aa}. 
\end{proof}

\section{An example: Obfuscated exponentiation \AppFrom{from \Sec{ss1055}}}\label{s1157Z}\label{s1357A}
In this section \C{give the details of the leakage calculations for the example from \Sec{ss1055}. For convenience we repeat the program at \Fig{f1143R}.}

\begin{figure}
{\Pf\begin{tabular}{ll}
\multicolumn{2}{l}{\Cx// B for base, the cleartext; E for exponent, the key:\ precondition is B,E >= 0,0 .} \\
\multicolumn{2}{l}{\Cx// P for power, the ciphertext.} \\
P:= 1 \\
\multicolumn{2}{l}{while E!=0\quad \Cx// Invariant is P*(B\textasciicircum E) = $b^e$, where $b,e$ are initial values of B,E .} \\
~~ D\From\ [2,3,5] & // D for divisor;\ uniform choice from \{2,3,5\}. \\
~~ R:= E mod D; & // R for remainder.\\
~~ if R!=0 then P:= P*B\textasciicircum R fi & // \fbox{Side-channel}:\ is E divisible exactly by D ? \\
~~ B:= B\textasciicircum D & // D is small:\ assume no side-channel here. \\
~~ E:= E div D & // State update of E here.\ (No side-channel.) \\
end \\
\multicolumn{2}{l}{// Now P=$b^e$ and E=0:\ but what has an adversary learned about the initial $e$~?}
\end{tabular}}

\bigskip
{\small Although our state comprises {\Pf B},{\Pf E},{\Pf P},{\Pf D},{\Pf R} we concentrate only on the secrecy of {\Pf E}. \C{In particular, we are not trying to discover {\Pf B} or {\Pf P} in this case; and {\Pf D},{\Pf R} are of no external significance afterwards anyway.}}
\caption{Defence against side channel analysis in exponentiation: a repeat of \Fig{f1143}}\label{f1143R}
\end{figure}

The well-known fast method of computing $B^E$ from a base $B$ exponent $E$ is a to employ a ``divide and conquer'' method to avoid the exponential time taken to perform $E$ multiplications. But in the most straightforward implementations, side-channels can reveal whether the current value of the exponent was decremented or divided by two, effectively revealing the exponent's bits one-by-one from least- to most significant. \C{In cryptographic applications, this reveals the secret key} \cite{Walter02a}.

A defence against that is to vary unpredictably the divisor that decreases the exponent on each iteration: not always 2, but say sometimes 3 and sometimes 5. This makes the algorithm (slightly) less efficient but, in compensation, much complicates the side-channel analysis that in the 2-case reveals everything.
The code in \Fig{f1143R} shows how: a fresh random divisor ${\Pf D}$ is chosen from a fixed set $\CalD$, independently each time round the loop; in \Fig{f1143R} we use $\CalD = \{2, 3, 5\}$. (The original, insecure algorithm uses $\CalD{=}\{2\}$.) Since a higher divisor offers more security, but takes longer to compute, the precise choice of $\CalD$ can be thought of as selecting a trade-off between information leakage and performance.%
\footnote{Walter \cite{Walter02a} gives a nice summary of various issues offering tradeoffs between leakage and performance. That algorithm is intended to provide more security by obfuscating further the pattern of if-branching. For simplicity we do not do that full analysis here.} 

We assume that all variables are secret: for the example we assume the adversary seeks the exponent ${\Pf E}$ only, that is the secret key, not e.g.\ a particular message. His side-channel is the {\Pf if}-statement: where he learns whether ${\Pf E}$ is exactly divible by ${\Pf D}$; complicating his analysis however is that he does not know exactly what ${\Pf D}$ is on each occasion.

We model this as an \HMM\ as follows.
We represent the state as a tuple $({\Pf B},{\Pf E},{\Pf P},{\Pf D},{\Pf R})$ containing the values of each variable in the program.
We can then model the statement ${\Pf D\From\ [2,3,5]}$ as a Markov update $M_D$ that uniformly assigns a fresh value to ${\Pf D}$ while leaving the rest of the variables untouched. That is, it maps each tuple $({\Pf B},{\Pf E},{\Pf P},{\Pf D},{\Pf R})$ to a tuple $({\Pf B},{\Pf E},{\Pf P},{\Pf D}',{\Pf R})$ with probability $\nicefrac{1}{|\CalD|}$ for each ${\Pf D}'\in \CalD$. 
The assignment then following can be modeled as a deterministic Markov $M_R$, mapping $({\Pf B},{\Pf E},{\Pf P},{\Pf D},{\Pf R})$ to $({\Pf B},{\Pf E},{\Pf P},{\Pf D},{\Pf E}\ {\Pf mod}\ {\Pf D})$.
We assume pessimistically that the observer can learn whether or not the statement ${\Pf P:= P*B\textasciicircum R}$ was executed, which logically is equivalent to observing the evaluation of the guard ${\Pf R!=0}$ of the ${\Pf if}$ statement. We model this flow as a
an \HMM\ comprising a deterministic channel $C$ and a Markov $M_P$. $C$ outputs, for each state $({\Pf B},{\Pf E},{\Pf P},{\Pf D},{\Pf R})$, whether ${\Pf R}$ is 0 or not. 
The (also deterministic) Markov part $M_P$ updates the state $({\Pf B},{\Pf E},{\Pf P},{\Pf D},{\Pf R})$ to $({\Pf B},{\Pf E},{\Pf P}{\Pf B}^{\Pf R},{\Pf D},{\Pf R})$ only when ${\Pf R}{\ne}0$, otherwise the state remains unchanged. 
Finally, the last two assignments correspond to deterministic Markov updates $M_B$ and $M_E$, mapping a state $({\Pf B},{\Pf E},{\Pf P},{\Pf D},{\Pf R})$ to $({\Pf B}^{\Pf D},{\Pf E},{\Pf P},{\Pf D},{\Pf R})$ and $({\Pf B},{\Pf E}\ {\Pf div}\ {\Pf D},{\Pf P},{\Pf D},{\Pf R})$ respectively.

We can express the program of \Fig{f1143} as an \HMM\ built as the sequential composition 
\[
\CM{}{M_D};\CM{}{M_R};\CM{C}{M_P};\CM{}{M_B};\CM{}{M_E};\quad \CM{}{M_D};\CM{}{M_R};\CM{C}{M_P};\ \cdots\hspace{2em},
\]
using the \HMM-step notations from \Sec{s1735}, whose good behaviour in representing our intentions is justified in \App{a1005}.

However, the exact number of iterations depends on the initial value of ${\Pf E}$. To simplify the analysis we will always assume  a fixed number of iterations (equal to the minimum number of iterations required for the program to yield the right result in every case). For instance, if {\Pf E} can take values from 0 to 15 (4 bits), then the maximum number of iterations is 4: on each iteration, {\Pf E} is divided at least by 2, and therefore after 4 iterations the value of {\Pf E} is guaranteed to be 0. (``Extra iterations'', i.e.\ when {\Pf E} is already zero, have no effect.) Once we have an \HMM, say $H$, we can calculate $\chan.H$ using \Lem{l1029} and, once we have $\chan.H$, we can calculate its min-leakage. From \Thm{t1203},
\Af{Do you mean  \Thm{t1203}? \Nx Yes, although it is a *bound* on the leakage (we still don't know neither the
correlation nor the gain function!)}
that gives also Dalenius leakage, which in this case represents the leakage related to the initial value of ${\Pf E}$. \Tbl{table:leakage} shows the result of computing this leakage \Tr{, in addition to the execution time of each case.}{for different sets of divisors.}

\section{Some notations and properties of \HMM\ steps \AppFrom{used in \App{s1157Z}}}\label{a1005}

In \Sec{s1735} it was remarked that the single, uniform definition of $({;})$ for \HMM's ``essentially'' treats pure markovs and pure channels differently. That difference is possible because markovs and channels are encoded differently within \HMM's. First we give the details for classical \HMM's; then further below we address abstract \HMM's.\label{a1320}
\Cf{Found this text ``orphaned'' in the appendix: where is it from?}
\label{g1255}
The \emph{degenerate} channel that leaks nothing about $\CalX$ can be regarded as having an anonymous singleton set as its observations, a single-column matrix containing only 1's. We write it $\NC_\CalX$ for ``no channel'' on state-space $\CalX$; usually we omit the $\CalX$. The degenerate markov that makes no change is the identity (matrix) on $\CalX$, which we write $\Id_\CalX$. We abbreviate $\CM{\NC}{M}$ by $\CM{}{M}$ and $\CM{C}{\Id}$ by $\CM{C}{}$, calling them (as \HMM's) \emph{pure} channels and pure markovs, and we note these properties of composition:
\begin{enumerate}
\item $\CM{C}{};\CM{}{M} = \CM{C}{M}$~, i.e.\ that $\CM{C}{M}$ is equivalent to ``just $C$'' and then ``just $M$''. 
\item\label{i1452} $\CM{C^1}{};\CM{C^2}{} = \CM{C^1{\Par}C^2}{}$ where $(C^1{\Par}C^2)_{x,(y^1,y^2)}$ defined $C^1_{x,y^1}C^2_{x,y^2}$ is the \emph{parallel composition} of its constituents, i.e.\ that ``$C^1$ and then $C^2$'' is the same as $C^1$ and $C^2$ together because $C^1$ passes the state on, unchanged, to $C^2$. 
\item $\CM{}{M^1};\CM{}{M^2} = \CM{}{M^1{\MMult}M^2}$, where $M^1{\MMult}M^2$ is the (usual) Markov composition of its constituents, i.e.\ matrix multiplication.
\end{enumerate}
The \HMM-composition of pure channels and composition of pure markovs is what allows us to combine channels and markovs and then treat them with the same definition of sequential composition, as is done in \App{s1157Z} with one long (associative) sequential composition of the loop body's constituents, then unfolded as explained there to eliminate the loop: their different purposes are automatically respected.

The essential property referred to in \Def{d1305} of \Sec{s1449} guarantees that this same respect for markovs' difference from channels is achieved in the semantic domain as well. Using \cite{McIver:2014ab,McIver:15} this can be exploited to remain entirely in the semantic domain, including the loop explicitly without having to unfold it.

\newpage
\section{Proofs and materials supporting \Sec{s1138}}\label{a1215}
{\Cx
\subsection*{On ``healthiness'' of abstract \HMM's\AppFrom{from \Sec{s1449}}}
The denotations of our $H$'s in $\CalX{\MFun}\CalY{\times}\CalX$ are functions in $\Dist\CalX{\Fun}\Dist^2(\CalX{\times}\CalX')$; but not all functions in that space are $\HMMSem{H}$ for some $H$. The ones that are have\Td{ here} been called abstract \HMM's, and they have properties that we use in the proofs below.

A common technique in semantics is to identify \emph{characteristic} properties of denotations, which can then lead to further insights. They are expressed in semantic terms, as we did for example in \cite[VIIA]{McIver:15} for collateral-\emph{unaware} abstract \HMM's, calling them ``healthiness conditions''.%
\footnote{\Cx A conspicuous example of this is Dijkstra's healthiness conditions for predicate transformers on sequential programs: the principal one was conjunctivity. Originally \cite{Dijkstra:76} Dijkstra proved by structural induction over program texts that all program denotations were healthy; later researchers instead used healthiness to characterise a subspace of predicate transformers, and then proved that all programs' denotations lay within it --- the same proof, but a different point of view. That more semantic view allowed later the expansion of healthiness, to include e.g.\ miracles and angelic choice, which in turn lead to new programming-language features that could denote the members of the expanded set.}
An advantage of that is the possible discovery of more general properties that secure computations' (semantics) should have.
Although we do not introduce collateral-aware healthiness here, it is a clear target for future work.
}

\subsection*{Proofs of lemmas and theorems\AppFrom{from \Sec{s1449}}}

\Cf{
Removed this (for now):\begin{quote}
\subsection*{Collateral-aware abstract \HMM's\AppFrom{from \Sec{s1449}}}
\C{We say that a function in our semantic space, some $h{\In}\Dist\CalX{\to}\Dist^2(\CalX{\times}\CalX')$, is \emph{healthy}} if for every collateral type $\CalZ$, there exists a \emph{unique} $\CalZ$-lifting \C{function} $\DaleniusVar{h}{\CalZ}$ such that the diagram in \Fig{f1041} commutes for every \Cr{channel}{right conditional} $Z{\In}\CalZ{\MFunR}\CalX$. That is, the healthiness of $h$ postulates the existence \C{for every collateral type $\CalZ$} of an $\DaleniusVar{h}{\CalZ}$ that depends only on $h$.
\Cd{This map is independent of the channel $Z$.}\C{This confuses me, at least for now :-) Which map? Is it $h^\CalZ$? By ``map'' do you mean ``function''? \T{Yes, I meant $h^\CalZ$ is independent of the right conditional $Z$. I meant to emphasises that we mean $\exists h^\CalZ\forall Z\dots$ rather than $\forall Z\exists h^\CalZ\dots$. From habit, I assume "function = one valued" and "map = total function".}}

Notice when the map $\DaleniusVar{h}{\CalZ}$ exists then it is necessarily unique because every joint distribution $\joint{\In}\Dist(\CalZ{\times}\CalX)$ can be decomposed into a right \Cr{channel}{conditional} $Z{\In}\CalZ{\MFunR}\CalX$ and right marginal $\pi{\In}\Dist\CalX$ such that $Z{\ApplyR}\pi = \joint$.%
\footnote{\Cx If $\pi$ is not full support on $\CalX$ then $Z$ is not unique: but see \Lem{l1241} below.}.
More precisely, if \C{some other} $h'{\In}\Dist(\CalZ{\times}\CalX)\C{{\Fun}\Dist^2(\CalZ{\times}\CalX')}$ made the diagram in \Fig{f1041} commute (\C{for every right conditional $Z\In\CalZ{\MFunR}\CalX$}),
then
\C{This seems strangely presented. I guess by ``exists'' you mean ``is single valued'', i.e.\ ``is a function in spite of the left hand arrow's going the wrong way?\T{Yes. That is another way of looking at it, i.e. we have a family of $h^Z$ (thus relational) instead of a single $h^\CalZ$. Notice however that the diagram implicitly assumes that every arrow is indeed an arrow, in particular, they are single-valued functions (they also need to be continuous and superlinear but we don't care about that, these have been proven in LiCS15).}}
\begin{equation}\label{e1150}
h'.\joint \Wide{=} \Dist(Z{\MMult}).\C{(}h.\pi\C{)} \Wide{=} \DaleniusVar{h}{\CalZ}.\joint\quad.
\end{equation}

Healthy $h$'s exist in abundance: if $H{\In}\CalX{\MFun}\CalY$ is a \HMM\ matrix, then for example $h=\HMMSem{H}$ is always healthy\C{, because we can define} $\DaleniusVar{\HMMSem{H}}{\CalZ}.\Pi = \Dist(Z{\MMult}).[\pi{\Apply}H]$, for every joint distribution $\Pi{\In}\Dist(\CalZ{\times}\CalX)$ with right channel resp. marginal $Z$ and $\pi$. The following lemma \C{shows that $\HMMSem{H}{\CalZ}$ does not depend on the particular choice of right conditional $Z$, where more than one is possible}.

\C{Lemma~\ref{l1241Z} $\HMMSem{H}{\CalZ}$ does not depend on the particular choice of right conditional $Z$, where more than one is possible.}

\begin{lemma}\label{l1241Z}
Let $H$ be a \HMM\ matrix, and $Z^{1,2}{\In}\CalZ{\MFunR}\CalX$ be channels and $\pi{\In}\Dist\CalX$ a distribution. If $Z^1{\ApplyR}\pi = \Pi = Z^2{\ApplyR}\pi$ then $\Dist(Z^1{\MMult}).[\pi{\Apply}H] = \Dist(Z^2{\MMult}).[\pi{\Apply}H]$.

\begin{proof}
This follows immediately from the fact that for every inner $\delta{\In}\Dist(\CalX{\times}\CalX')$ of \C{the hyper} $[\pi{\Apply}H]$, we have $Z^1{\MMult}\delta = Z^2{\MMult}\delta$ when $Z^1{\ApplyR}\pi = Z^2{\ApplyR}\pi$. \C{In particular, when $\pi_x{=}0$ because $\Pi_{z,x}{=}0$ for all $z$, which is the only case where $Z^1.x$ and $Z^2.x$ can differ, then also $\delta_x{=}0$ and so $(Z^1{\MMult}\delta)_{z,x'}=(Z^2{\MMult}\delta)_{z,x'}=0$ for all $z,x'$.}
\C{Need to think about this more.}
\end{proof}
\end{lemma}

\end{quote}
}
{\Cx
Lemma~\ref{l1241} shows that $h^\CalZ$ in \Fig{f1041} is well defined whenever $h{=}\HMMSem{H}$ for some $H$. The principal fact is that possible variation in the choice of the right-conditional $Z$, at left, does not affect the value produced by $\Dist(Z\MMult)$ at right.
\begin{lemma}\label{l1241}
Let $h$ be an abstract \HMM; then $h^\CalZ$ from \Fig{f1041} is well defined. \AppFrom{from \Sec{s1449}} \\ 

\begin{proof} Let $h{=}\HMMSem{H}$ for some $H\In\CalX{\MFun}\CalY{\times}\CalX$, and  let $\Pi$ be a joint distribution in $\Dist(\CalZ{\times}\CalX)$ with right marginal $\pi\In\Dist\CalX$ and $Z\In\CalZ{\MFunR}\CalX$ a right-conditional so that $\Pi{=}Z{\ApplyR}\pi$. If $Z$ is unique, then the result is trivial because $h^\CalZ.\Pi$ is given by the composition of the bottom- and right-hand arrows acting on $\pi$ in \Fig{f1041}.

If however $Z$ is not unique, it can only be at elements $x$ where $\pi_x$ is 0, the well known issue that the all-zero column $\Pi_{-,x}$ cannot be normalised: but equally well known is that ``it doesn't matter'' because any arbitrary value chosen for $Z_{-,x}$ will be multiplied by 0 in $\Pi{=}Z{\ApplyR}\pi$.

It doesn't matter on the right, either, because a property of abstract $h$'s inherited from their originating $H$'s is that if $\pi.x{=}0$ then the sub-vector $(\pi{\Apply}H)_{x,-,-}$ of the joint distribution $\pi{\Apply}H$ is all zero; that means, in turn, that in all inners $\delta$ of the hyper $[\pi{\Apply}H]$ derived from that joint distribution (by abstracting from $y$) the probability $\delta_{x,x'}$ for that $x$, and for for any $x'$, will be zero as well. Thus, again, variation of $Z$ at that $x$ does not matter.

{\Tx In all cases,
\begin{equation}\label{e1150}
\DaleniusVar{h}{\CalZ}.\joint \Wide{=} \Dist(Z{\MMult}).\C{(}h.\pi\C{)}
\end{equation}
holds whenever $\Pi{=}Z{\ApplyR}\pi$.}
\end{proof}
\end{lemma}
}

\Cf{I suggest leaving the below out: we want abstract \HMM's to be the healthy $h$'s, I think; and we don't want to focus people on pathological cases.
\begin{quote}\Bx
There are also many unhealthy abstract \HMM s. For instance, the existence of $\HMMSem{H}^\CalZ$ requires that $H$ ``copies'' the initial value into the first component in $\CalX{\times}\CalX'$. Without this condition, \Lem{l1241} does not hold.\end{quote}}

\subsection*{Proof for \Thm{t0639C} \AppFrom{from \Sec{s1449}}}

Since $h^{{\times}\CalZ} = h^{\CalZ^2{\times}\CalX}{\circ}\Dist\zeta$, it suffices to prove the following result.

\begin{theorem}
	Let $h_{1,2}$ be two abstract \HMM's. Then $h_1\Ref h_2$ iff $\DaleniusVar{h_1}{\CalZ} \Ref \DaleniusVar{h_2}{\CalZ}$ for all extension $\CalZ$.
	
\begin{proof}	
	Let us assume that $h_1\Ref h_2$ and let $\Pi\In\Dist(\CalZ{\times}\CalX)$ with a right-conditional $Z$ and right marginal $\pi$, i.e. $Z{\ApplyR}\pi = \Pi$. Let $\Delta_i = h_i.\pi$ for $i=1,2$. Recall from \cite[Def.~6]{McIver:12} that two hypers satisfy $\Delta_{1}{\Ref}\Delta_{2}$ in $\Dist^2\CalX^2$ iff there exists a super $\Gamma\In\Dist^3\CalX^2$ such that $\Delta_1 = \mu_{\Dist\CalX^2}.\Gamma$ (the outer average) and $\Delta_2 = \Dist\mu_{\CalX^2}.\Gamma$ (the inner average)
	\footnote{Here, $(\Dist,\mu,\LHyp{-})$ is the usual Giry monad, see \cite{Giry:81,McIver:12,McIver:15}.}. We say that $\Gamma$ \emph{witnesses} the refinement $\Delta_{1}{\Ref}\Delta_{2}$.
	Let us show that the super  $\Dist^2(Z{\MMult}).\Gamma\In\Dist^3(\CalZ{\times}\CalX)$ witnesses the refinement $\Dist(Z{\MMult}){.\Delta_1}\Ref\Dist(Z{\MMult}){.\Delta_2}$ in $\Dist(\CalZ{\times}\CalX)$, i.e.
	\[
	\Dist(Z{\MMult}){.\Delta_1}\Wide{=}\mu_{\Dist(\CalZ{\times}\CalX)}.\left(\Dist^2(Z{\MMult}){.\Gamma}\right) \qquad\textrm {and }\qquad \Dist (Z{\MMult}){.\Delta_2}\Wide{=} \Dist\mu_{\CalZ{\times}\CalX}.\left(\Dist^2 (Z{\MMult}){.\Gamma}\right)~.
	\]
	
	Both equalities can be stated using the diagrams given in \Fig{f1132A}-(a,b). Let us show that these diagrams indeed commute.
	\begin{figure}[!h]
		\centering
		\begin{minipage}{.32\textwidth}
			\begin{tikzcd}
				\Dist^2\CalX^2\arrow{r}{\Dist (Z{\MMult})}& \Dist^2(\CalZ{\times}\CalX) \\
				\Dist^3\CalX^2\arrow[swap]{r}{\Dist^2 (Z{\MMult})}\arrow{u}{\Dist\mu_{\CalX^2}}&\arrow[swap]{u}{\Dist\mu_{(\CalZ{\times}\CalX)}} \Dist^3(\CalZ{\times}\CalX) \\
			\end{tikzcd}
			\\\centering (a)
		\end{minipage}
		\begin{minipage}{.32\textwidth}
			\begin{tikzcd}
				\Dist^2\CalX^2\arrow{r}{\Dist (Z{\MMult})}& \Dist^2(\CalZ{\times}\CalX) \\
				\Dist^3\CalX^2\arrow[swap]{r}{\Dist^2 (Z{\MMult})}\arrow{u}{\mu_{\Dist\CalX^2}}&\arrow[swap]{u}{\mu_{\Dist(\CalZ{\times}\CalX)}} \Dist^3(\CalZ{\times}\CalX) \\
			\end{tikzcd}
			\\\centering (b)
		\end{minipage}
		\begin{minipage}{.32\textwidth}
			\begin{tikzcd}
				\Dist\CalX^2\arrow{r}{(Z{\MMult})}& \Dist(\CalZ{\times}\CalX) \\
				\Dist^2\CalX^2\arrow[swap]{r}{\Dist (Z{\MMult})}\arrow{u}{\mu_{\CalX^2}}&\arrow[swap]{u}{\mu_{\CalZ{\times}\CalX}} \Dist^2(\CalZ{\times}\CalX) \\
			\end{tikzcd}
			\\\centering (c)
		\end{minipage}
		\caption{The multiplication $(Z{\MMult})$ commutes with inner ($\Dist\mu$) and outer ($\mu_{\Dist}$) averages.}\label{f1132A}
	\end{figure}
	
	For the diagram in \Fig{f1132A}-(a)(inner-average), it suffices to show \Fig{f1132A}-(c) commutes 	because $\Dist$-lifting preserves commutative diagrams. This essentially follows from the linearity of $(Z{\MMult})$. Let us first recall the definition of multiplication in the Giry monad. Given a (polish) space $\cal S$ and a Borel set $O{\subseteq} \cal S$, the multiplication $\mu_{\cal S}$ is defined by the expected value of the evaluation function $\epsilon_O{\In}\Dist{\cal S}{\to} \Real$ where $\epsilon_O.\delta = \delta.O$ for every $\delta{\In}\Dist{\cal S}$. We have, for every $\Delta{\In}\Dist^2{\cal S}$,
	\begin{equation}\label{e2044}
		(\mu_{\cal S}.\Delta).O = \int \epsilon_O\mathrm{d}\Delta~.
	\end{equation}
	We simply write $\epsilon_s$ when $O$ is the singleton $\{s\}$. Recall also that given a measurable function $f{\In}\cal R{\to}\cal S$, the push-forward of $f$ by $\Dist$ is $\Dist f{\In}\Dist\cal R{\to}\Dist\cal S$ where, for every Borel set $O{\subseteq}\cal S$ and measure $\delta{\In}\Dist\cal R$, we have
	\begin{equation}\label{e2155}
		\Dist f.\delta.O \Wide{=}\delta.(f^{-1}.O)\Wide{=}\delta.\{r\ |\ f(r){\in}O\}~.
	\end{equation}
	A very useful result links the constructor $\Dist$ with integrals. Let $f{\In}\cal R{\to}\cal S$ and $g{\In}\cal S{\to}\Real$ be measurable functions and $\delta{\In}\Dist\cal R$ be a Borel measure. We know that $\Dist f.\delta{\in}\Dist\cal S$ so we can integrate $g$ wrt it. In fact, Giry has proven in \cite[Sec. 3 p.70]{Giry:81} that 
	\begin{equation}\label{e1502}
		\int g\mathrm{d}\Dist f.\delta\Wide{ = }\int g{\circ}f\mathrm{d}\delta~.
	\end{equation}
	We are now ready to prove that all diagrams in \Fig{f1132A} commute. For  \Fig{f1132A}-(c), let $\Delta\In\Dist^2\CalX^2$ be an arbitrary hyper and $(z,x'){\In}\CalZ{\times}\CalX$. On the one hand,	
	\begin{Reason}
		\Step{}{
			\left(\mu_{\CalZ{\times}\CalX}{\circ}\Dist(Z{\MMult}).\Delta\right)_{z,x'}
		}
		\StepR{$=$}{Eqn.~\Eqn{e2044}}{
			\int \epsilon_{(z,x')}\mathrm{d}\Dist(Z{\MMult}).\Delta
		}
		\StepR{$=$}{Eqn.~\Eqn{e1502}}{
			\int \epsilon_{(z,x')}{\circ} (Z{\MMult})\mathrm{d}\Delta
		}
	\end{Reason}
	On the other hand, we have
	\begin{Reason}
		\Step{}{
			(Z{\MMult}(\mu_{\CalX^2}.\Delta))_{z,x'}
		}	
		\StepR{$=$}{Def.~matrix multiplication}{
			\sum_x Z_{z,x}(\mu_{\CalX^2}.\Delta)_{x,x'}
		}
		\StepR{$=$}{Eqn.~\Eqn{e2044}}{
			\sum_x Z_{z,x}\int \epsilon_{(x,x')}	\mathrm{d}\Delta
		}
		\StepR{$=$}{Integration is linear}{
			\int\sum_x Z_{z,x}\epsilon_{(x,x')}\mathrm{d}\Delta
		}
	\end{Reason}
	Thus, it remains to show that the integrands in the last lines of the reasoning above are the same. Let $\delta{\In}\Dist{\CalX^2}$, we have
	\[
		\left(\epsilon_{(z,x')}{\circ(Z{\MMult})}\right).\delta \Wide{=} (Z{\MMult}\delta)_{z,x'}\Wide{=} \sum_x Z_{z,x}\delta_{x,x'}\Wide{=}\left(\sum_x Z_{z,x}\epsilon_{(x,x')}\right).\delta
	\]
	Hence, \Fig{f1132A}-(c) commutes and its $\Dist$-lift gives the inner-average commutative diagram.
	
	We now prove that \Fig{f1132A}-(b) (outer-average) commutes. Let $\Gamma{\In}\Dist^3\CalX^2$ and $O$ be a Borel subset of $\Dist^2(\CalZ{\times}\CalX)$. On the one hand, we have 
	\begin{Reason}
		\Step{}{
			\left(\mu_{\Dist(\CalZ{\times}\CalX)}{\circ}\Dist^2(Z{\MMult})\right).\Gamma.O
		}
		\StepR{}{Def.~($\circ$)}{
			\left(\mu_{\Dist(\CalZ{\times}\CalX)}.\left(\Dist^2(Z{\MMult}).\Gamma\right)\right).O
		}
		\StepR{$=$}{Eqn.~\Eqn{e2044}}{
			\int \epsilon_O\mathrm{d}\Dist^2(Z{\MMult}).\Gamma
		}
		\StepR{$=$}{Eqn.~\Eqn{e1502}}{
			\int \epsilon_O{\circ}\Dist(Z{\MMult})\mathrm{d}\Gamma
		}
	\end{Reason}
	
	On the other hand, we have
	\begin{Reason}
		\Step{}{
			\left(\Dist(Z{\MMult}){\circ}\mu_{\Dist\CalX^2}\right).\Gamma.O
		}
		\StepR{$=$}{Def.~($\circ$)}{
			\Dist(Z{\MMult}).(\mu_{\Dist\CalX^2}.\Gamma).O
		}
		\StepR{$=$}{Def.~push-forward \Eqn{e2155}}{
			(\mu_{\Dist\CalX^2}.\Gamma).((Z{\MMult})^{-1}.O)
		}
		\StepR{$=$}{Eqn.~\Eqn{e2044}}{
			\int \epsilon_{(Z{\MMult})^{-1}.O}\mathrm{d}\Gamma
		}
	\end{Reason}
	As before, it remains to show that the integrands are equal. For every hyper $\Delta{\In}\Dist\CalX^2$, we have
	\begin{Reason}
		\Step{}{
			\left(\epsilon_O{\circ\Dist(Z{\MMult})}\right).\Delta
		}
		\StepR{$=$}{Def.~($\circ$)}{
			\epsilon_O.(\Dist(Z{\MMult}).\Delta)
		}
		\StepR{$=$}{Def.~$\epsilon$}{
			\Dist(Z{\MMult}).\Delta.O
		}
		\StepR{$=$}{Def.~push-forward \Eqn{e2155}}{
			\Delta.((Z{\MMult})^{-1}.O)
		}
		\StepR{$=$}{Def.~$\epsilon$}{
			\epsilon_{(Z{\MMult})^{-1}.O}.\Delta	
		}
	\end{Reason}
	This concludes that \Fig{f1132A}-(b) commutes.
	
	Conversely, assume $\DaleniusVar{h_1}{\CalZ}{\Ref}\DaleniusVar{h_2}{\CalZ}$ for every extension $\CalZ$. For $\CalZ = \CalX$, we have $\DaleniusVar{h_1}{\CalX}{\Ref}\DaleniusVar{h_2}{\CalX}$. Let $\pi\In\Dist\CalX$ and $\Pi{\In}\Dist\CalX^2$ such that $\Pi$ has $\pi$ on the diagonal and $0$ anywhere else. It follows from Equation \Eqn{e1150} that $h_i.\pi = \DaleniusVar{h_i}{\CalX}.\Pi$ and thus $h_1{\Ref}h_2$.
\end{proof}
\end{theorem}

\subsection*{Proof that $\HMMSem{-}$ preserves sequential composition \AppFrom{from \Sec{s1449}}}
\begin{lemma}\label{l1653}
Let $H^{1,2}$ be \HMM\ matrices. We have $\HMMSem{H^1\GComp H^2} = \HMMSem{H^1}
\C{\BSemi}
\DaleniusVar{\HMMSem{H^2}}{\CalX}$.

\begin{proof}
Let $\pi{\In}\Dist\CalX$ be a prior. On the one hand, let $y_{1,2}$ be observations associated to the inner $\delta$ in $\HMMSem{H^1\GComp H^2}.\pi$. That is, for pair $x,x'{\in}\CalX$ of initial and final value:

\[
	\delta_{x,x'} \Wide{=}\frac{\pi_{x}(H^1\GComp H^2)_{x,(y_1,y_2),x'}}{\Norm_{H^1\GComp H^2,\pi,(y_1,y_2)}} 
\]
where the normalisation constant is defined as in \Sec{s1540}:
\begin{equation}\label{e1540}
	\Norm_{H,\pi,y} = \sum_{x,x'}\pi_{x} H_{x,y,x'}~.
\end{equation}
On the other hand, let us have a look at the inners of $\HMMSem{H^1}\BSemi \DaleniusVar{\HMMSem{H^2}}{\CalX}.\pi$. Let one such inner be $\delta\In\Dist(\CalX{\times}\CalX')$. By Equation~\Eqn{e1150}, we have $\DaleniusVar{\HMMSem{H^2}}{\CalX}.\gamma = \Dist(Z{\MMult}).\HMMSem{H^2}.\RMarg{\gamma}$ where the right-conditional and right marginal satisfy $Z{\ApplyR}\RMarg{\gamma} = \gamma$. Therefore, there exists some inner $\gamma{\In}\Dist(\CalX{\times}\CalX')$ of $\HMMSem{H^1}.\pi$ such that $\delta = Z{\MMult}\rho$ for some inner $\rho{\In}\Dist\CalX{\times}{\CalX'}$ of $\HMMSem{H^2}.\RMarg{\gamma}$ and $Z{\ApplyR}\RMarg{\gamma} = \gamma$. Thus
\[
	\gamma_{x,u}\Wide{=}\frac{\pi_{x} H^1_{x,y_1,u}}{\Norm_{H^1,\pi,y_1}}
	\qquad\textrm{and}\qquad
	\rho_{u,x'} \Wide{=}\frac{\RMarg{\gamma}_u H^2_{u,y_2,x'}}{\Norm_{H^2,\RMarg{\gamma},y_2}} 
\]
for some observations $y_{1,2}$. Hence
\begin{Reason}
\Step{}{
	\delta_{x,x'}
}
\StepR{$=$}{$\delta = Z{\MMult}\rho$}{
	(Z{\MMult}\rho)_{x,x'}
}
\StepR{$=$}{Def.~$Z{\MMult}$}{
	\sum_u Z_{x,u}\rho_{u,x'}
}
\StepR{$=$}{Def.~of $\rho$ above}{
	\sum_u Z_{x,u}\frac{\RMarg{\gamma}_u H^2_{u,y_2,x'}}{\Norm_{H^2,\RMarg{\gamma},y_2}}
}
\StepR{$=$}{$Z{\ApplyR}\RMarg{\gamma} = \gamma$}{
	\sum_u \gamma_{x,u}\frac{ H^2_{u,y_2,x'}}{\Norm_{H^2,\RMarg{\gamma},y_2}}
}
\StepR{$=$}{Def.~of $\gamma$ above}{
	\sum_u \frac{\pi_{x} H^1_{x,y_1,u}}{\Norm_{H^1,\pi,y_1}}\frac{ H^2_{u,y_2,x'}}{\Norm_{H^2,\RMarg{\gamma},y_2}}
}
\StepR{$=$}{Def.~$H^1;H^2$}{
	\frac{ \pi_{x} (H^1;H^2)_{x,(y_1,y_2),x'}}{\Norm_{H^1,\pi,y_1}\Norm_{H^2,\RMarg{\gamma},y_2}}
}
\end{Reason}
It remains to prove that $\Norm_{H^1\GComp H^2,\pi,(y_1,y_2)} = \Norm_{H^1,\pi,y_1}\Norm_{H^2,\RMarg{\gamma},y_2}$ (this essentially tells us that it does not matter whether we apply the normalisation process for each component in the sequential composition or we leave it until the end and just carry out normal \HMM\ multiplication at the matrix level).
\begin{Reason}
\Step{}{\Norm_{H^1,\pi,y_1}\Norm_{H^2,\RMarg{\gamma},y_2}}
\StepR{$=$}{Def.~$\Norm_{H^2,\RMarg{\gamma},y'}$}{
	\sum_{u,x'}{\RMarg{\gamma}_u H^2_{u,y_2,x'} \Norm_{H^1,\pi,y_1}}
}
\StepR{$=$}{$\RMarg{\gamma}_u = \sum_x\gamma_{x,u}$}{
	\sum_{u,x'}{\sum_x\gamma_{x,u} H^2_{u,y_2,x'} \Norm_{H^1,\pi,y_1}}
}
\StepR{$=$}{Substituting $\gamma_{x,u}$}{
	\sum_{u,x'}{\sum_x\frac{\pi_{x} H^1_{x,y_1,u}}{\Norm_{H^1,\pi,y_1}} H^2_{u,y_2,x'} \Norm_{H^1,\pi,y_1}}
}
\StepR{$=$}{Simplification with $\Norm_{H^1,\pi,y_1}$ ($\dagger$)}{
	\sum_{u,x'}{\sum_x\pi_{x} H^1_{x,y_1,u} H^2_{u,y_2,x'}}
}
\StepR{$=$}{Arith.}{
	\sum_{x,x'}\pi_{x} \sum_u{H^1_{x,y_1,u} H^2_{u,y_2,x'}}
}
\StepR{$=$}{Def.~$(H^1\GComp H^2)_{{x},(y_1,y_2),x'}$}{
	\sum_{x,x'}\pi_{x} (H^1\GComp H^2)_{x,y_2,x'}
}
\StepR{$=$}{Def.~of $\Norm_{(H^1\GComp H^2),\pi,(y_1,y_2)}$}{
	\Norm_{(H^1\GComp H^2),\pi,(y_1,y_2)}~.
}
\end{Reason}
($\dagger$) This step assumes the normalising constant $\Norm_{H^1,\pi,y_1}$ is not zero. If it is zero, then the equality we want to prove clearly holds.

Hence, $\HMMSem{H^1\GComp H^2}.\pi$  and  $(\HMMSem{H^1}\BSemi \HMMSem{H^2}).\pi$ have the exact same inners associated to the same respective outer probabilities. 
\end{proof}
\end{lemma}

\subsection*{Proof that sequential composition \AppFrom{from \Sec{s1449}} \\ respects behavioural equivalence}

\begin{theorem}\label{t1329}
	If $H \Ref H'$ and $K \Ref K'$ then $H\GComp K \Ref H'\GComp K'$.
\end{theorem}

Let us firstly prove a very important property of \HMM s: they transform gain functions. We construct, for every $g{\In}\GainF{\cal W}\CalX^2$, a gain function $g^H{\In}\GainF{\CalY{\to}{\cal W}}\CalX^2$ where
\begin{equation}
g^H.\sigma.(x,x') \Wide{=} \sum_{y,u} g.(\sigma.y).(x,u)H_{x',y,u}~,
\end{equation}
for every strategy $\sigma{\In}\CalY\to\CalW$ and $x,x_0{\in}\CalX$. We have the following properties of $g^H$.

\begin{lemma}\label{l1449}
Let $H,K$ be \HMM s and $\pi{\In\Dist\CalX}$ be a prior. For every gain function $g{\In}\GainF{\CalY{\to}\CalY{\to}{\cal W}}\CalX^2$, we have:
\begin{enumerate}
	\item $\GTest{\LHyp{\LRJoint \pi H}}{g} = V_{g^H}\LHyp{\pi}$,\label{p1449a}
	\item $g^{H;K} = (g^{H})^{K}$.\label{p1449b}
\end{enumerate} 

\begin{proof}
For \ref{p1449a}., we have 
\begin{Reason}
	\StepR{}{Def.~$V_g[\pi{\Apply}H]$}{
		\sum_y\max_{w{\In}{\cal W}} \sum_{x,x'}g.w.(x,x') H_{x,y,x'}\pi_{x}
	}
	\StepR{$=$}{Swap $\max$ and $\sum_y$}{
		\max_{\sigma{\In}\CalY\to{\cal W}}\sum_{y}\sum_{x,x'}g.(\sigma.y).(x,x') H_{x,y,x'}\pi_{x}
	}
	\StepR{$=$}{Def.~$g^H$}{
		\max_{\sigma{\In}\CalY\to{\cal W}}\sum_{x}g^H.\sigma.(x,x)\pi_{x}
	}
	\StepR{$=$}{Def.~$\GTest{\LHyp{\pi}}{g}$}{
		\GTest{\LHyp{\pi}}{g^H}
	}	
\end{Reason}
For \ref{p1449b}., let $\sigma{\In}\CalY\to\CalY\to\CalW$. We have
\begin{Reason}
	\Step{}{(g^K)^H.\sigma.(x,x')}
	\StepR{$=$}{Def.~$g^H$}{
		\sum_{y,u}g^K.(\sigma.y).(x,u) H_{x',y,u}
	}
	\StepR{$=$}{Def.~$g^K$}{
		\sum_{y,u} \sum_{y',v}g.(\sigma.y.y').(x,v) K_{u,y,v} H_{x',y,u}
	}
	\StepR{$=$}{Swap sums}{
		\sum_{y} \sum_{y',v}g.(\sigma.y.y').(x,v) \sum_uH_{x',y,u} K_{u,y,v}
	}
	\StepR{$=$}{Def.~$H;K$}{
		\sum_{y,y'}\sum_v g.(\sigma.y.y').(x,v) (H;K)_{x',(y,y'),v}
	}
	\StepR{$=$}{Def.~$g^{H;K}$}{
		g^{H;K}.\sigma.(x,x')
	}
\end{Reason}
as required.	
\end{proof}
\end{lemma}
Now, we can prove \Thm{t1329}.

\begin{proof}[Proof of \Thm{t1329}]
Let $H\Ref H'$, $K\Ref K'$, $g$ be a gain function and $\pi\In\Dist\CalX$ be a prior. On the one hand, assume $H\Ref H'$ and let us show that $H;K\Ref H';K$.
	
\begin{Reason}
	\Step{}{
		\GTest{\LHyp{\LRJoint{\pi}{(H\GComp K)}}}{g}	
	}
	
	\StepR{$=$}{\Lem{l1449}.\ref{p1449a}}{
		\GTest{\LHyp{\pi}}{g^{H;K}}
	}
	
	\StepR{$=$}{\Lem{l1449}.\ref{p1449b}}{
		\GTest{\LHyp{\pi}}{(g^K)^H}
	}
	
	\StepR{$=$}{\Lem{l1449}.\ref{p1449a}}{
		\GTest{\LHyp{\LRJoint \pi H}}{g^K}
	}
	
	\StepR{$\geq$}{$H\Ref H'$}{
		\GTest{\LHyp{\LRJoint \pi H'}}{g^{K}}
	}
	
	\StepR{$=$}{\Lem{l1449}.\ref{p1449a} and \Lem{l1449}.\ref{p1449b}}{
		\GTest{\LHyp{\LRJoint{\pi}{(H'\GComp K)}}}{g}		
	}
\end{Reason}

On the other hand, assume $K\Ref K'$ and let us show that $H';K\Ref H';K'$. For each observation $y$ of $H'$, we consider the associated inner $\delta^y{\In}\Dist\CalX^2$ and outer $p^y{\In}[0,1]$ where
\[
(\delta^y)_{x,x'} \Wide{=}\frac{\pi_xH'_{x,y,x'}}{\Norm_{H',y,\pi}}
\]
where $\Norm_{H',y,\pi} = \sum_{x,x'}\pi_xH'_{x,y,x'}$ is the usual column normalising constant. Then
\begin{Reason}
	\Step{}{
		\GTest{\LHyp{\LRJoint{\pi}{(H'\GComp K)}}}{g}
	}
	
	\StepR{$=$}{Def.~$\delta^y$'s}{
		\sum_y p^y\GTest{\LHyp{\LRJoint{\delta_y}{K}}}{g}
	}
	
	\StepR{$\geq$}{$K\Ref K'$}{
		\sum_y p^y\GTest{\LHyp{\LRJoint{\delta_y}{K'}}}{g}
	}
	
	\StepR{$=$}{Def.~$\delta^y$'s}{
		\GTest{\LHyp{\LRJoint{\pi}{(H'\GComp K')}}}{g}
	}
\end{Reason}
Hence, $H;K\Ref H';K'$ follows by transitivity of $\Ref$.
\end{proof}

\newpage
\section{Reduction to Bayes vulnerability via a collateral context \AppFrom{from \Sec{s0908}}}\label{a1002}

In \Sec{s0908} we asked whether the refinement relation $\Ref$ from \Sec{s1330} was too strong, and recalled that in \cite{mcivermeinicke10a} a construction was given that reduced refinement to Bayes vulnerability, at least for closed systems without collateral correlations: if $P{\NRef}Q$ then there was a context ``post-compose with program $R$\/'' and a prior $\pi$ such that the two hypers
\[
  \Delta^P~=~\HMMSem{P;R}.\pi \WideRm{and}  \Delta^Q~=~\HMMSem{Q;R}.\pi
\]
had \emph{Bayes} vulnerabilities in particular ``the wrong way around'', i.e.\ that $V_\BVg(\Delta^P)<V_\BVg(\Delta^Q)$. That justified the failure of refinement, for if $P{\Ref}Q$ then we cannot, for any prior and $R$, have that $Q;R$ is \emph{more} Bayes vulnerable than $P;R$.

That argument does not however work directly for refinements that are rejected because of \emph{collateral} correlations: the post-composing $R$ does not have access to the initial states of $P,Q$ if those programs modify the state. A very direct (and somewhat brutal) work-around would be to re-use the construction of \cite{mcivermeinicke10a} by putting $P,Q$ into a context that preserved the input by first copying it into {\Pf Z} say, then executing $P,Q$ resp.\ and finally introducing an $R$ as before but one that operated not on {\Pf X} directly but rather on the preserved copy held in {\Pf Z}. Because neither of $P,Q$ assign to {\Pf Z}, it would still contain {\Pf X}'s original value; but \emph{leaks} in $P,Q$ from {\Pf X} would be reflected in different hypers $\Dist^2(\CalZ{\times}\CalX)$ resulting from each. That is, informally we would have a collateral context containing declaration {\Pf var Z{:}$\CalX$} and the code
\[
 {\Pf Z:= X;}\quad P;\quad R_{\Pf Z}  \hspace{4em}\textrm{and}\hspace{4em} {\Pf Z:= X;}\quad Q;\quad R_{\Pf Z} ~,
\]
where $R_{\Pf Z}$ is the distinguishing $R$ from \cite{mcivermeinicke10a} but operating on {\Pf Z} instead of {\Pf X}.

A more convincing approach however is to use the gain function $g$ that distinguishes $P,Q$ to make a ``pre'' collateral correlation $\Pi^g$ instead of a ``post'' assignment $R^g$. It has been shown that the $\Pi^g$ can be derived from the $R^g$; but a more direct route is the following.

Assume the programs $P,Q$ are indistinguishable wrt.\ the hypers on their final states (else we could simply appeal to \cite{mcivermeinicke10a}), but that they differ wrt.\ their information-flow effect on the initial state. Using  $\chan$ from \Lem{l1029} in \Sec{s1101}, take $\chan.P$ and $\chan.Q$, the effective channels of $P,Q$ resp.\ and let $\Delta^P,\Delta^Q$ resp.\ be the hypers resulting from the action of those channels respectively on the prior $\pi$. By assumption $\Delta^P{\NRef}\Delta^Q$, and so there is some gain-function $g$ with $V_g(\Delta^P)<V_g(\Delta^Q)$. It is shown in \cite{Smith:aa}, but considering only channels (not \HMM-programs), that there is indeed a collateral correlation $\Pi^g\In\Dist(\CalW{\times}\CalX)$ such that 
\[
 V_\BVg[\Pi^g\MMult\chan.P] \Wide{<}  V_\BVg[\Pi^g\MMult\chan.Q] ~,
\]
where $[\Pi^g\MMult\chan.P]$ is the hyper resulting from the joint distribution in $\Dist(\CalW{\times}\CalY)$ formed by the matrix multiplication of the joint distribution $\Pi^g$ by the stochastic $\chan.P$.

Thus we have shown that if the $g$-vulnerabilities of $\chan.P$ and $\chan.Q$ wrt.\ some $g\In\CalW{\Fun}\CalX{\Fun}\Real$ and $\pi\In\Dist\CalX$ mandate $P{\NRef}Q$, then the collateral \emph{Bayes} vulnerabilities of $P,Q$ are the wrong way around for some collateral correlation $\Pi^g$ (whose right-marginal is of course $\pi$); general $g$-discsimination is again, as in \cite{mcivermeinicke10a}, reduced to Bayes-vulnerability discrimination and compositionality.

\end{document}